\documentclass[pdflatex,sn-mathphys-num]{sn-jnl}

\usepackage{graphicx}%
\usepackage{multirow}%
\usepackage{amsmath,amssymb,amsfonts}%
\usepackage{amsthm}%
\usepackage{mathrsfs}%
\usepackage[title]{appendix}%
\usepackage{xcolor}%
\usepackage{textcomp}%
\usepackage{manyfoot}%
\usepackage{booktabs}%
\usepackage{algorithm}%
\usepackage{algorithmicx}%
\usepackage{algpseudocode}%
\usepackage{graphicx}%
\usepackage{caption}%
\usepackage{subcaption}%
\usepackage{listings}%
\usepackage{anyfontsize}%

\usepackage[version=4,arrows=pgf-filled,
textfontname=sffamily,
mathfontname=mathsf]{mhchem}
\newcommand{\oxides}{29}
\newcommand{\metals}{38}

\newcommand{\Cu}{128,296}
\newcommand{\Ir}{55,756}

\newcommand{\adsorbates}{41}

\newcommand{\reacenergy}{$E_{r}$}
\newcommand{\reacbarrier}{$E_{a}$}

\newcommand{\screensubsampleError}{0.15}
\newcommand{\screentotalsamples}{90,781}
\newcommand{\screenconvergedsamples}{90,181}
\newcommand{\screentestcases}{2,115}

\theoremstyle{thmstyleone}%
\theoremstyle{thmstyletwo}%

\theoremstyle{thmstylethree}%

\begin{document}

\title[Article Title]{Systematic Fine-Tuning of MACE Interatomic Potentials for Catalysis}

\author*[1]{\fnm{Nima} \sur{Karimitari}}\email{karimitn@mailbox.sc.edu}

\author[2]{\fnm{Jacob} \sur{Clary}}

\author[2]{\fnm{Derek} \sur{Vigil-Fowler}}

\author[3]{\fnm{Ravishankar} \sur{Sundararaman}}

\author[4,5]{\fnm{G\'abor} \sur{Cs\'anyi}}

\author*[1,6]{\fnm{Christopher} \sur{Sutton}}\email{ca.sutton@utoronto.ca}

\affil[1]{\orgdiv{Department of Chemistry and Biochemistry}, \orgname{University of South Carolina}, \orgaddress{ \city{Columbia}, \postcode{29208}, \state{SC}, \country{United States}}}

\affil[2]{\orgdiv{Materials, Chemical, and Computational Science Directorate}, \orgname{National Laboratory of the Rockies}, \orgaddress{ \city{Golden}, \postcode{80401}, \state{CO}, \country{United States}}}

\affil[3]{\orgdiv{Department of Materials Science and Engineering}, \orgname{Rensselaer Polytechnic Institute}, \orgaddress{ \city{Troy}, \postcode{12180}, \state{NY}, \country{United States}}}

\affil[4]{\orgdiv{Engineering Laboratory}, \orgname{University of Cambridge}, \orgaddress{\street{Trumpington St}, \city{Cambridge}, \postcode{CB21PZ}, \country{United Kingdom}}}

\affil[5]{\orgdiv{Max Planck Institute for Polymer Research}, \orgaddress{\street{Ackermannweg 10}, \city{Mainz}, \postcode{55128}, \country{Germany}}}


\affil[6]{\orgdiv{Department of Materials Science and Engineering}, \orgname{University of Toronto}, \orgaddress{\street{184 College Street}, \city{Toronto}, \postcode{M5S 3E4}, \state{ON}, \country{Canada}}}

\abstract{Once trained, machine-learned interatomic potentials (MLIPs) provide a fast and accurate method to gain insight into catalytic reaction pathways. Key to the performance of MLIPs is how their training set is constructed. 
In this work, we compared the performance of 9 MLIPs with different training sets and training strategies, both from-scratch (FS) and fine-tuning (FT) large foundation models. In order to examine how these MLIPs perform for realistic reactions, we then tested these models on the both the reaction energies (\reacenergy{}) and the reaction energy barriers (\reacbarrier{}) for a diverse set of 141 chemical reactions such as the \ce{CO2} reduction reaction (\ce{CO2RR}) to \ce{C2} and \ce{C3} products, propane dehydrogenation to propylene, hydrogen intercalation on palladium, as well as an out of distribution reaction of oxygen evolution reaction (OER) on metal-oxides. 
Our results show that FS models with 5 \% to 10 \% of perturbed (high-energy) configurations from MD or contour exploring (CE), can reduce the error by over 2× compared to models trained only on relaxation trajectories.
In contrast, FT MLIPs are less sensitive to training set sampling and perform well on out-of-distribution reactions. For example, an MLIP fine-tuned on metallic catalysts achieves an MAE of 0.30 eV for OER on iridium oxide polymorphs, which is 0.08 eV lower than out-of-box MACE-MH-1 and 0.14 eV lower than the best performing FS MLIP. Similarly, an MLIP fine-tuned to O and OH adsorption on metal oxides gives an \reacbarrier{} MAE of 0.19 eV for out-of-distribution \ce{CO2RR} reaction on Cu (which is equivalent to an FS model that was trained to in-distribution carbon bond-breaking events on Cu surfaces). 
Finally, a large MLIP fine-tuned to 49,860 configurations shows the best performance across both metallic and metal-oxide catalysts, which was used to screen a large-left out, \screentotalsamples{}, of bimetallic alloys, achieving an MAE of 0.15 eV for \reacenergy{}, even for adsorbates on unseen (e.g., 532) Miller index surfaces. Overall, this work identifies the necessary training set configurations for FS models and FT MLIPs for different catalyst types and reports an accurate MLIP that is useful broadly for catalytic reactions.}

\maketitle

\section{Introduction}\label{sec1}
Catalyzed chemical reactions are fundamental to the physical sciences and central to important fields like electrochemistry, fuels synthesis, and battery development.\cite{intro1,intro2,intro3}. Studying heterogeneous catalytic reactions requires understanding two key quantities: the reaction energy between adsorbed states (\reacenergy{}) and the activation barrier (\reacbarrier{}).

Ab initio \reacenergy{} calculations for heterogeneous catalysis typically rely on DFT optimization of adsorbate configurations on catalyst surfaces.
This approach is manageable for simple adsorbates and surfaces but can become computationally prohibitive for more complex catalyst-adsorbate combinations that exponentially increase the combinations of binding sites, defect states, surface disorder, adsorbate orientations/coverages, and interactions with the gas/liquid phase. 

Predicting \reacbarrier{} via transition state searches is particularly challenging, as it requires accurate knowledge of the PES near both local minima and saddle points. The nudged elastic band (NEB) method\cite{ci-neb} has been widely used to identify transition states, but has an cost that scales with number of intermediate images and can converge slowly for poor initial structure guesses. The dimer\cite{dimer_method} and growing string\cite{growing_String_method} methods can also help locate unknown transition state structures\cite{dimer_method} and improve robustness for highly nonlinear reaction coordinates\cite{growing_String_method}. Although each of these methods can be efficient for specific reactions, the overall workflow bottleneck is the computational cost of DFT.

Machine learning (ML) can help reduce this bottleneck by bypassing the need for running DFT and have been used to accelerate adsorption energy predictions across diverse catalytic surfaces\cite{binding_1,binding_2,binding_3}. While these studies show promising performance for targeted reactions, common challenges are the cost of generating electronic features (e.g. $d$-band center) or using chemically intuitive guesses of adsorption sites.

With the emergence of MLIPs as a promising tool to predict energy and forces that scales linearly with the number of atoms, one can probe the PES near local minima or saddle points of any catalytic reaction using hundreds to thousands of atoms\cite{mlip-guide} with a speed up of up to $10^{6}$ faster than DFT with comparable accuracy. They also offer improved flexibility and reactivity over classical molecular dynamics (MD) force fields. Recent equivariant message passing MLIP architectures can efficiently learn atom environment representations and have greatly improved accuracy and transferability over previous approaches \cite{MPNNGilmer2017, Batzner2022:nequip, Batatia2022ThePotentials}. For heterogeneous catalysis, the reliance on large datasets to train MLIPs such as the Open Catalyst datasets \cite{oc20} "out of the box" might be a limitation, as performance remains heavily dependent on data availability and quality, which may restrict their generalizability when determining both \reacenergy{} and \reacbarrier{} 


Wood, et al. \cite{wood2025umafamilyuniversalmodels} have used a mixture of experts model to train a family of universal models (UMA) on a combined dataset comprised of OC22 (oxide electrocatalysts) \cite{oc22}, OC25 (solid-liquid interfaces) \cite{sahoo2025opencatalyst2025oc25}, OMOL (molecules) \cite{levine2025openmolecules2025omol25}, OpenDac \cite{sriram2025opendac2025dataset} and OMAT (bulk inorganic materials)\cite{barrosoluque2024openmaterials2024omat24}and is able to estimate \reacenergy{} within 0.1 eV of DFT, for 74 \% of \reacenergy{} reported in AdsorbML task. In another effort, Batatia, et al. pretrained a base model to OMAT dataset \cite{barrosoluque2024openmaterials2024omat24}, then fine-tuned it using a multi-head replay with 10\% of the OMAT to avoid catastrophic forgetting, and 6 other heads that consists of SPICE \cite{spice}, RGD1 \cite{RGD1}, Material project \cite{MP}, OC20 \cite{oc20}, OMOL \cite{levine2025openmolecules2025omol25}, and MATPES R2SCAN \cite{Kuner2025}. This model is shown to have good performance for adsorbate-surface interaction prediction, with an MAE of 0.095 eV for the S24 \cite{mp0} and 0.138 eV for OC20 benchmark tests. The CatTSunami project evaluated whether MLIPs could predict $\Delta$\reacbarrier{} relative to DFT with an accuracy of least 0.1 eV. They found that the MLIP-predicted NEBs were successful for 56\% of the 900 reactions studied, which improved to 86\% when the $\Delta$\reacbarrier{} were calculated using single-point DFT energies at the MLIP's initial, transition, and final state structures. Although these studies show that MLIPs can successfully be used for calculating \reacbarrier{} and \reacenergy{} of heterogeneous catalytic systems, it remains unclear which components of large-scale training datasets are most relevant for improving model accuracy.


When training from-scratch (FS) MLIPs for specific catalytic applications, it often involves multiple iterations of sampling (mostly using the uncertainty of a committee of MLIPs) of configurations generated uniformly, randomly or through some additional technique such as minima-hopping \cite{minimahop}, molecular dynamics (MD), or along an NEB trajectory. For example, Schaaf, et al. \cite{neb_strategy} trained an accurate MLIP (MAE of 0.05 eV for \reacbarrier{}) for \ce{CO2} hydrogenation to methanol on \ce{In2O3} in six active-learning loops with configurations taken initially from bulk structures and  bare surfaces, and additional configurations with higher energy uncertainty taken from geometry relaxations, MD trajectories, and NEB pathways. With the availability of large foundation models and recent fine-tuning methods, including multi-head replay, one can achieve similar or better accuracy and more generalizability than FS models \cite{mp0,batatia2025crosslearningelectronicstructure}. However, it remains unclear how sensitive these fine-tuned (FT) models are to training set diversity and generation via different sampling techniques, and how choices in training set generation manifest in \reacenergy{} and \reacbarrier{} prediction accuracy for heterogeneous catalysis. 

In this work, we first highlight the importance of having multiple cycles of retraining with diverse configurations taken from MD or contour exploration in predicting \reacenergy{} and \reacbarrier{}. Then, we show fine-tuning from MACE-MH-1 provides a more efficient way of achieving even more accurate MLIPs, often with smaller datasets than FS models. We demonstrate cross-learning between metallic and metal-oxide catalysts and how bond-breaking events on a metallic catalyst in the training set can improve the model performance for OER steps on metal-oxides polymorphs. Similarly, models fine-tuned to datasets consisting of only metal-oxides have decent performance when applied to NEB calculations on metallic catalysts. In the last section, we used our largest FT MLIP to predict \screentotalsamples{} adsorption energies of unseen adsorbates on binary transition metal alloys with a high accuracy.

\section{Methods}
\subsection{DFT calculations}\label{sec:dft}
All DFT calculations used Quantum Espresso (QE) \cite{Giannozzi_2009}, ultrasoft pseudopotentials (GBRV) \cite{GBRV}, and the BEEF-vdW functional\cite{beef}, which accounts for van der Waals interactions crucial for catalysis. The plane-wave and charge density cutoffs were set at 500 eV and 5000 eV, respectively. The k-point grids were generated using Monkhorst-Pack \cite{MP} with a density of 50 k-points per \AA{} for in-plane lattice vectors, along with a 0.15 eV Fermi smearing, and at least 12 \AA{} of vacuum. Spin-polarized calculations were performed for systems containing transition metals with partially occupied $d$-orbitals. Initial magnetic moments of 3, 3, 2, 1  were chosen for Fe, Mn, Co and Ni, respectively. 

\subsection{Reaction energy and NEB calculations}\label{sec:neb}
Structures were optimized for all reaction energy predictions using a pre-conditioned-LBFGS \cite{preLBFGS} algorithm until the forces predicted by the MLIP on all atoms were less than $5~\text{meV/\AA}$. All structures used for the first and last NEB images were relaxed with either DFT or MLIP calculators prior to NEB calculations. For the test set case studies for \reacbarrier{}, we used the initial and final steps as provided in their studies, but fully relaxed them with the MLIP until all forces were less than $5~\text{meV/\AA}$. We ignored all the previously calculated DFT intermediate images for the NEB pathways and instead used the image-dependent pair potential (IDPP) method \cite{idpp} to generate intermediate images between the initial and final MLIP optimized structures. Due to the substantially reduced inference cost of the MLIP for NEBs, we doubled the number of intermediate images as compared to the previously reported DFT pathways to improve reaction pathway smoothness. It should be noted that using additional intermediate images (up to 10$\times$) can further improve the smoothness if needed. A NEB calculation was considered converged once the normal forces to images were all less than 50 $meV$/\AA.

\subsection{Contour exploration}\label{sec:ce}

Contour exploring (CE) was used to improve training data sampling for non-equilibrium structures. Conceptually, CE attempts to follow the same potential energy contour across the PES rather than follow gradients downhill to a local minimum, with the details of the approach discussed by Waters and Rondinelli \cite{Waters_2021}. Briefly, by using the Taylor expansion of the Frenet-Serret formulas, one can update atomic positions via displacements, $\Delta r$, calculated using the curvature of the potential energy contour: $\Delta r= \Delta r_{\parallel} + \Delta r_{\perp}  + \Delta r_{drift}$. Here, $\Delta r_{\perp} $ ensures that the total energy reaches the target value and act similar to a thermostat in MD simulations (and can also referred to as a potentiostat $U_{target}$), $\Delta r_{\parallel}$ is the motion along the potential energy contour calculated at every step using the arc length and magnitude of the curvature, and $\Delta r_{drift}$ ensures that the trajectory does not get stuck in a loop. It is important to note that at every step of the trajectory, the total velocity is maintained to be perpendicular to the total force. Therefore, the tangential vector ($T$) also remains perpendicular to the forces. 

\subsection{MACE training}\label{sec:mace}
We utilized the MACE framework, an equivariant message passing tensor network, for constructing the MLIP \cite{MACE2022,Batatia2025DesignSpace}. MACE offers state-of-the-art accuracy and has been successfully trained to large datasets, including the Materials Project, with benchmark accuracy~\cite{mp0}. The MACE architecture uses many-body equivariant message passing 
 and parameterize a mapping from atomic coordinates and atomic numbers to the total
potential energy by decomposing it into atomic energies.

The MACE loss function is
\begin{equation}
\begin{aligned}
\mathcal{L} = &\frac{\lambda_E}{B} \sum_{b=1}^{B} \left(\frac{ \hat{E}_b - E_b }{N_b}\right)^2 \\
& + \frac{\lambda_F}{3B} \sum_{b=1}^{N} \sum_{i_b=1,\alpha=1}^{N_b,3} \left( - \frac{\partial \hat{E}_b}{\partial \mathbf{r}_{i_b,\alpha}} - F_{i_b,\alpha} \right)^2,
\end{aligned}
\end{equation}
where $B$ is the number of batches, $N_b$ is the number of atoms in the batch, $E_b$ is the DFT energy, $\hat{E}_b$ is the predicted energy, $F_{i_b,\alpha}$ is the DFT force component in the direction $\alpha$ of atom $i_b$. $i_b$ denotes the index within batch $b$. $\lambda_F$ and $\lambda_E$ are the weights of the model,

For the FS models, the $\lambda_F$ and $\lambda_E$ are set to 10 and 1000, respectively, in the first 400 epochs and then switched to 1000 and 1000, respectively, for the last 100 epochs. The specifications of the MACE models used in this work are given in Table \ref{table:model_specs}.

\begin{table}[]
\caption{Specifications of the MACE models used in this work.}
\label{table:model_specs}
\begin{tabular}{@{}cc@{}}
\toprule
\textbf{Model parameter} & \textbf{Value} \\ \midrule
Number of chemical channels & 512 \\
Maximum equivariance order, L & 1 \\
Single layer cutoff radius (\AA) & 6 \\
Number of layers & 2 \\ \bottomrule
\end{tabular}
\end{table}

For FT models, we started from the checkpoint of the foundational MACE-MH-1 model, and used only the OMAT head for fine-tuning purposes. To avoid catastrophic forgetting during the fine-tuning, we used the multi-head replay approach, which has been shown to be promising in other studies \cite{mp0,batatia2025crosslearningelectronicstructure}. In particular, we used one head from the MACE-MH-1 OMAT dataset with 30K configurations taken from the overall OMAT dataset (energy and force labels were taken from MACE-MH-1 predictions), and the other head as a fine-tuning dataset from this work. Each of FT models were trained independently, and those that shared a training set also had the same splitting of training and validation sets. Fine-tuning was performed for a total of 120 epochs, with energy and forces weights of 1 and 10 for the first 80 epochs and then switched to 10 and 10, respectively, for the rest of the fine-tuning. We used a learning rate of 0.0001, two orders of magnitude smaller than what we used for the FS models. All other hyperparameters are the same as the original MACE-MH-1 model.

\subsection{Molecular dynamics and uncertainty analysis}\label{sec:md}
MD simulations were performed using the atomic simulation environment (ASE) package \cite{ase} in the NVT ensemble at 300 K using a Nosé–Hoover thermostat \cite{nose,hoover}. For reaction energy uncertainty analysis, MD simulations were performed using a committee of 3 MLIPs trained to the same data sets but initialized with different random seeds. The relative force uncertainty for each atom among the 3 models, $f_{rel}^i$, is defined as
\begin{equation}
f_{rel}^i=\frac{\sigma_i}{\left|\bar{F}_i\right|+\epsilon},
\label{eq:rel_force_uncert}
\end{equation}
where $\sigma_i$ and $\bar{F}_i$ denote the standard deviation and mean of forces of the model committee members on atom \textit{i}, and $\epsilon$ is a regularizer set to 0.2 eV/\AA{} to avoid diverging ratios for small forces. MD simulations were terminated if $max(f_{rel}^i) < 0.2$ after a step. 

\subsection{Reaction energy predictions}
All $E_r$ were calculated using $E_{r}=E_{products}-E_{reactants}$, where $E_{reactants}$ and $E_{products}$ are the total energy of all reactants and products, including any gas phase reference energies, respectively. For each reaction, all reactant or product structures containing a surface are optimized with both DFT ($E_{r}^{DFT}$) and an MLIP ($E_{r}^{MLIP}$) independently, starting from the provided structure in each dataset. However, given that each MLIP model is primarily trained to surface reactions, we still used DFT values for gas phase reference energies. These energies were calculated using the same DFT functional and pseudopotentials as the surface calculations. All $E_a$ were calculated using $E_{a}=E_{transition}-E_{reactants}$, where $E_{transition}$ is the total energy of the highest energy image along an NEB pathway.


\section{Datasets}\label{sec:datasets}

We divided the overall set of training data into 3 groupings: relaxation trajectories, MD sampling, and CE trajectories, with each grouping anticipated to provide different types of chemical information to the MLIP. All the configurations for the training set are initially taken from 6 studies and supplemented with additional calculations. Table \ref{table:training} summarizes training set details such as the number of unique surface compositions, reactions, facets, and total configurations, and we refer the reader to the original papers for additional details. Similarly, the same statistics for the the final test set application datasets are shown in Table \ref{table:test}. 

\begin{figure}
    \centering  
\includegraphics[width=\textwidth]{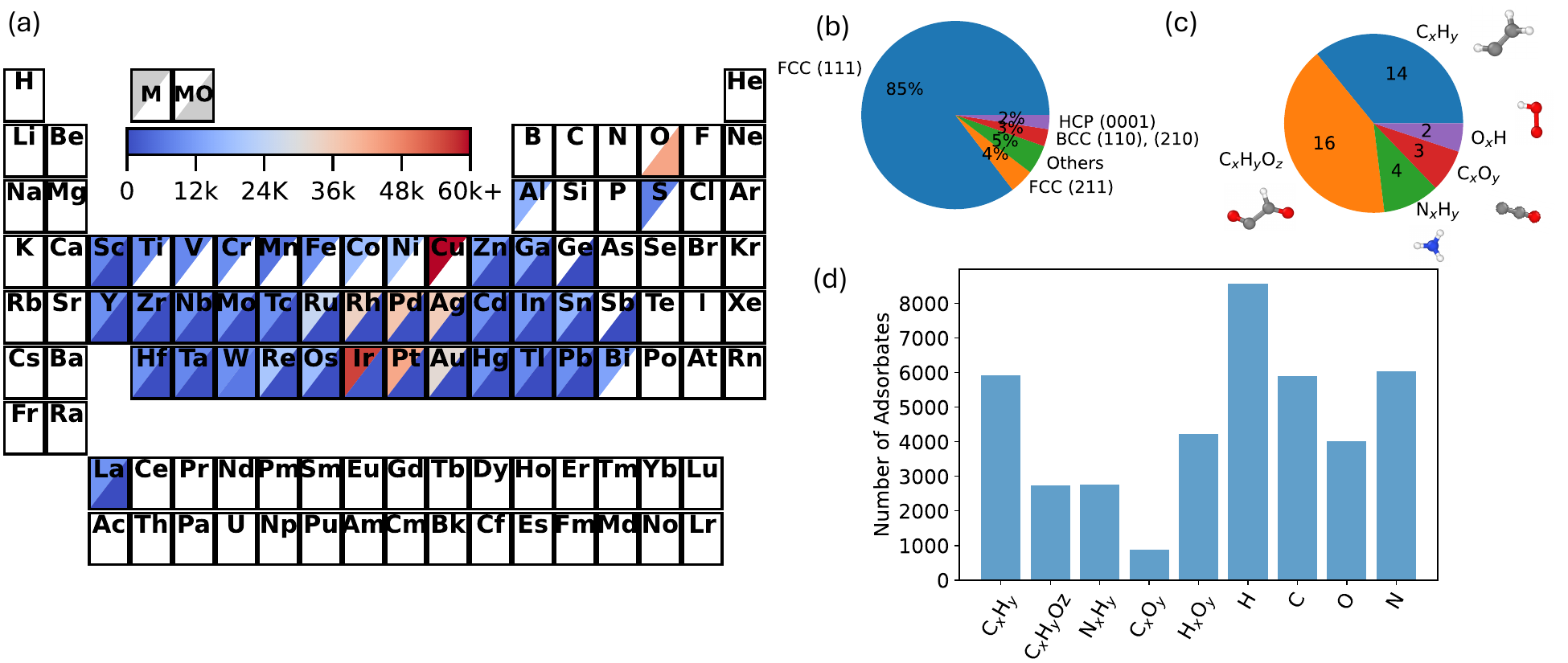}
    \caption{(a) The total number of elements, excluding the adsorbates, in all the configurations of the training dataset across metallic (M) and metal-oxides (MO) catalysts. (b) The percentages of the types of surface termination in the training set. (c) Unique types of organic molecules in the training set, and (d) total number of occurrence of each type of adsorbates in the training set.} 
    \label{fig:dataset} 
\end{figure}

\begin{itemize}
    \item \underline{\textit{Relaxation Trajectory Configurations}}: The bimetallic alloy (BMA) surface dataset was constructed by randomly selecting 20,000 configurations from 200,000 previously published chemisorbed single-atom/diatomic adsorbates on 2,035 bimetallic alloy surfaces with 5 different stoichiometric ratios.\cite{bma} The selection was only constrained such that the final training set would have an equal distribution of adsorbates (H, C, N, O, and OH). Only the first (highest energy state) and last (fully relaxed) structures along the relaxation trajectory for each system were selected rather than the entire trajectory to improve model knowledge of both local minima and non-equilibrium bonding configurations without excessive data duplication. The distribution of all atoms in this training set is shown in Fig. S1.
    \item  \underline{\textit{Metallic MD Configurations}}: We performed MD simulations for 10 ps in the NVT ensemble at $T=300~K$ on the reported optimized structures from 4 studies, gas conversion to \ce{C2+} oxygenates on FCC (111) transition metals \cite{schumann}, ammonia synthesis \cite{Wangachieveing, montoya}, and dehydrogenation of ethane \cite{HANSEN2019161} using the FS-BMA model. Then configurations were selected from these trajectories with a spacing of at least 0.5 ps along each trajectory. In total, the selected configurations from these 4 datasets and MD trajectories is consisted of 76 reactions, 5 surface terminations, 192 surface compositions. 

    \item \underline{\textit{Metal-Oxide MD Configurations}}:
    We performed similar MD simulations on obtained structures of OH and H adsorption on metal oxides. \cite{comer} In total we added 2,209 metal-oxide configurations to our dataset. Fig. \ref{fig:dataset}a shows that this metal oxide set spans 29 metals, with Table \ref{table:training} providing further details.
    \item \underline{\textit{CE Configurations}}: To create bond-breaking configurations, we used CE by evolving the systems along potential energy contours \cite{Waters_2021}. As shown in SI Sec. S2, a major advantage of CE over double-ended NEB calculations is that it does not require both the reactant and product states to be known to generate new training data. The potentionostat parameter, $U_{target}$, is adjustable and enables the creation of multiple rare events for the same starting structure, and does so with substantially fewer ionic steps as compared to MD simulations. We performed CE starting from the same structures used to generate the Metallic MD Configurations above (Table~\ref{table:training}), with the potentiostat set at 150 meV/atom higher than the energy of the MLIP optimized structure (i.e., the local minima) with $U_{target}=150~\text{meV/atom}$ and gathered a total of 2,209 configurations. 
\end{itemize}

An overall summary of the entire training set is shown in Figure \ref{fig:dataset}. Broadly, it contains information about elements across the periodic table, with a particular emphasis on metal (M) catalysts (\metals{} elements) and metal-oxide (MO) (\oxides{} TM elements) catalysts, and organic adsorbates (\adsorbates{} adsorbates). The upper triangles in Figure~\ref{fig:dataset}a show the amount of training data for the metal surfaces while the lower triangles show the amount of training data for the metal-oxide surfaces. Elements with partially filled 3$d$ shells (e.g., Cu, Fe, Ni) were excluded from the MO training set because these systems generally require special care in quantum chemical calculations, such as the use of DFT Hubbard $U$ corrections or quasiparticle self-consistent GW approximation calculations, which would introduce DFT setting inconsistency across all of our datasets. Metallic Cu (\Cu{} occurrences) and Ir (\Ir{} occurrences) atoms are the most dominant elements in the dataset due to their importance in the \ce{CO2RR} and the OER that we focus on here for targeted applications. Figure~\ref{fig:dataset}b shows the distribution of surface facets included, highlighting the large amount of fcc (111) facet data. We note that the generation of additional DFT training data for higher indexed facets with undercoordinated surface atoms that could be present in catalytic nanoparticles is highly desirable. For the adsorbates, Figure~\ref{fig:dataset}c shows the distribution of organic adsorbates with variable stoichiometries, while Figure~\ref{fig:dataset}d shows the overall adsorbate distribution in the data.

\begin{table}[]
\caption{All types of datasets that are used in training the MLIPs. The first four rows correspond to the metallic catalyst datasets and the fifth row corresponds to the metal-oxide dataset. We note that for the BMA dataset in row 1, we directly take configurations from the relaxation trajectory, but for the rest of the datasets, we sampled using MD or CE simulations starting from the provided structures in each dataset. The number of unique reactions only considers the reactants and products of the reactions, and ignores any enumerations across surface terminations/compositions}
\label{table:training}
\addtolength{\tabcolsep}{-0.4em}
\begin{tabular}{@{}cccccc@{}}
\textbf{Dataset} & \textbf{Facets} & \textbf{\begin{tabular}[c]{@{}c@{}}Surface\\ comps.\end{tabular}} & \textbf{Configs.} & \textbf{\begin{tabular}[c]{@{}c@{}}Unique\\ Reactions\end{tabular}} & \textbf{\begin{tabular}[c]{@{}c@{}}All\\ Reactions\end{tabular}} \\ \midrule
Bimetallic TM alloys \cite{bma} & (111) & 1924 & 39,983 & 6 & - \\
C$_{2+}$ oxygenates \cite{schumann} & (111) & 23 & 2,056 & 25 & 170 \\
NH$_3$ synthesis \cite{Wangachieveing, montoya} & (111), (211), (110), bcc(110) & 71 & 2,577 & 6 & 512 \\
C$_2$H$_6$ dehydrogenation \cite{HANSEN2019161} & (111), hcp(0001) & 94 & 2,551 & 45 & 840 \\
O, OH adsorption \cite{comer} & (100), (110), (210) & 88 & 2,693 & 6 & 217 \\ \midrule
\textbf{Total} & \textbf{8} & \textbf{2200} & \textbf{49,860} & \textbf{88} & \textbf{1569}
\end{tabular}
\end{table}

\begin{table}[]
\caption{Aggregated details of the metallic and metal-oxide test sets for reaction energy predictions in Sec. \ref{sec:applications} showing the unique surface compositions, reactions, facets, total number of configurations that the MLIP is applied to, and the total number of reaction energies that are calculated.}
\label{table:test}
\begin{tabular}{@{}cccccc@{}}
\textbf{Dataset} & \textbf{\begin{tabular}[c]{@{}c@{}}Number\\ of facets\end{tabular}} & \textbf{\begin{tabular}[c]{@{}c@{}}Surface\\ compositions\end{tabular}} & \textbf{Configs.} & \textbf{\begin{tabular}[c]{@{}c@{}}Unique\\ Reactions\end{tabular}} & \textbf{\begin{tabular}[c]{@{}c@{}}All\\ Reactions\end{tabular}} \\ \midrule
Metallic catalysts \cite{test_set_clark,test_set_landers,test_set_li,test_set_peng_role,test_set_peng_trends_2022,test_set_saini_electronic_2022,test_set_snider,test_set_tang_modeling,test_set_tangfrom,test_set_tetteh,test_set_wang,test_set_yang} & 16 & 524 & 12,458 & 136 & 291 \\
Metal-oxide catalysts \cite{oxide_test_set_flores,oxide_test_set_lee,oxide_test_set_shilong,oxide_test_set_strickler} & 9 & 78 & 1,032 & 5 & 214
\end{tabular}
\end{table}
\section{Results}

\subsection{MLIP Training}\label{sec:dataset}

We first discuss the training strategy for three FS models, the training dataset, and their performance when applied to the test set applications in Table \ref{table:test_error}. Next, we focus on six FT models that were trained to different combinations of the above datasets and compare their performance in Table \ref{table:ft_test_error}. 
\subsubsection{From-scratch Training}

\begin{itemize}
    \item \underline{FS-BMA:\textit{ Relaxation trajectories}}: This model was trained to ca. 40K configurations directly taken from the BMA dataset, but has a mediocre performance and achieves an MAE of 0.659 eV when predicting the metal test set \reacenergy{}. Despite the site diversity provided to the FS-BMA model, it is not expected to perform well for reactions involving adsorbates with 3 or more atoms, such as intermediates in the \ce{CO2RR}, due to the dataset only containing H, C, N, O, and OH as adsorbates. Larger adsorbates typically have more binding configurations to the surface with more local minima. We note that this model fails to converge when applied to metal-oxide catalysts or NEBs for \reacbarrier{} predictions.
    
    \item \underline{FS-BMA+MD:\textit{ Addition of MD configurations}}: The FS-BMA model performance is improved for all three test applications by augmenting the training set with about 7,659 metal and metal-oxide configurations generated using MD. In particular, the model achieves an MAE of 0.310 eV for metallic catalyst \reacenergy{}, better than the performance of the unmodified large foundation models (UMA and MACE-MH-1), but performs poorly for \reacenergy{}(MO) with an MAE of 0.476 eV, higher than corresponding values for UMA (OMAT) or MACE-MH-1 (OMAT)
    
    \item \underline{FS-All:\textit{ Addition of CE configurations}}: To inform the FS models of bond-breaking chemistry, we further augmented the training dataset with 2,209 configurations generated using CE sampling using the FS-MD model. This model achieves an MAE of 0.175 eV for \reacbarrier{}(M) predictions, about half of the error calculated using the FS-BMA+MD model, within 0.05 eV of the UMA and MACE-MH-1 models' MAE. 
\end{itemize}




\begin{table}[]
\caption{Summary table of the total number of configurations used in each from-scratch training set (N$_\mathrm{FS{\text -}TS}$), MAE of \reacenergy{} and \reacbarrier{} in eV for the metallic (M) and metal-oxide (MO) test set studies using different versions of the MLIP trained in this work, and comparisons of our model performance with the UMA and MACE models. }
\label{table:test_error}
\begin{tabular}{@{}ccccc@{}}
\toprule
\multirow{2}{*}{\textbf{Model}} & \multirow{2}{*}{\textbf{N$_\mathrm{FS{\text -}TS}$}} & \multicolumn{3}{c}{\textbf{Datasets}} \\ \cmidrule(l){3-5} 
 &  & \textbf{$E_{r}$ (M)\cite{test_set_clark,test_set_landers,test_set_li,test_set_peng_role,test_set_peng_trends_2022,test_set_saini_electronic_2022,test_set_snider,test_set_tang_modeling,test_set_tangfrom,test_set_tetteh,test_set_wang,test_set_yang}} & \textbf{$E_{r}$ (MO) \cite{oxide_test_set_flores,oxide_test_set_lee,oxide_test_set_shilong,oxide_test_set_strickler}} & \textbf{$E_{a}$ (M) \cite{oxide_test_set_flores,oxide_test_set_lee,oxide_test_set_shilong,oxide_test_set_strickler}} \\ \midrule
FS-BMA & 39,584 & 0.659 & -- & -- \\
FS-BMA+MD & 47,651 & 0.310 & 0.476 & 0.398 \\
FS-All & 49,860 & 0.279 & 0.398 & 0.175 \\
\hline
UMA (OC20) & $\sim$500 M & 0.403 & 0.889 & 0.131 \\
UMA (OMAT) & $\sim$500 M & 0.558 & 0.243 & 0.453 \\
MACE-MP0 & $\sim$1 M & 0.745 & 0.466 & 0.493 \\
MACE-MH-1 (OC20) & $\sim$117 M & 0.578 & 0.556 & 0.176 \\
MACE-MH-1 (OMAT) & $\sim$117 M & 0.478 & 0.322 & 0.220 \\ \bottomrule
\end{tabular}
\end{table}

\subsubsection{Fine-tuning to MACE-MH-1}\label{sec:fine-tune}
Next, we use the datasets described in Sec. \ref{sec:dataset} and fine-tuned 6 different MLIPs, each with different set of training configurations. For the fine-tuning, we used a multi-head replay strategy, where one head is for our own dataset and the other is the OMAT head of MACE-MH-1. For each of these models, we used 30K configurations from OMAT, with labels of energy and forces taken from the MACE-MH-1 model and all hyper parameters kept the same across the models. Further details of the fine-tuning approach are discussed in Sec. \ref{sec:mace}. 

The model fine-tuned to the BMA dataset (FT-BMA) with just relaxation trajectories is better than best-performing from scratch model (FS-All) that was trained to relaxation trajectory and additional 9,877 MD and CE configurations In addition, with no metal oxide in the training set, the \reacenergy{}(MO) MAE is 0.334 eV, about 0.06 eV lower than a model that had about 2,692 metal-oxides configurations in the training set. Similarly, we next fine-tuned 3 additional models to only metallic MD configurations (FT-MD(M)), only metal-oxides MD configurations (FT-MD(MO)), or only CE configurations (FT-CE). The FT-MD(M) model performs worse than the FT-BMA model with a 0.15 eV increase in \reacenergy{}(M) MAE. This can be explained by the lack of bimetallic systems in the MD trajectories, but also suggests that the high energy and forces from MD configurations do not improve model predictions in this case, and that a more diverse set of chemical environments is a major factor in model performance. Another interesting observation is that the FT-MD(M) model MAE for \reacenergy{}(MO) is 0.428 eV while the FT-BMA MAE is 0.334 eV, despite both models lacking any metal-oxides in the training set. We attribute the improved FT-BMA performance to it learning information about the metal-oxide chemical environment from the higher number of bimetallic configurations used for training. Furthermore, the FT-CE model, which was fine-tuned using 1,789 Cu surface structures and 420 Ag, Ir, Pd, Pt, and Rh surface structures generated using CE, performs better for \reacenergy{}(MO) (MAE = 0.305 eV) than the FT-BMA and FT-MD(M) models. It also has almost identical performance to the FT-MD(MO) model (MAE = 0.306 eV), which was fine-tuned using only metal-oxides MD configurations. This result is likely due to O-H bond breaking events being present in the CE metallic catalyst dataset, which appears to improve O and OH adsorption predictions on metal-oxide catalyst, despite no metal-oxide catalysts being explicitly present in the training set.

Next, we expanded the BMA fine-tuning training set by adding metallic MD configurations, leading to a new model (FT-BMA+MD(M)) exhibiting an almost identical performance to the FT-BMA model. This is likely because the 30K configurations from OMAT already contain many high temperature MD structures, so adding more MD configurations does not provide much new chemical information to the model. Finally, after fine-tuning on all of our datasets (FT-All), the \reacenergy{}(M) and \reacbarrier{} (M) MAE are unchanged, the \reacenergy{}(MO) MAE decreases by 0.07 eV. This is because the previous FT models had no metal-oxides in the training set. These results further indicates that one does not need a diverse set of high energy and forces for fine-tuning, given that MACE-MH-1 has already been trained to a very diverse set of datasets and the 30K structures used from OMAT prevent the catastrophic forgetting of the MACE-MH-1 head during the fine-tuning.

\begin{table}[]
\caption{Summary of the MAE (in eV) for six ft-MLIPs (from MACE-MH-1) to predict \reacenergy{} and \reacbarrier{} for the metal (M) and metal-oxide (MO) datasets. N$_\mathrm{FT{\text -}TS}$ is the number of training configurations used in fine-tuning.}
\label{table:ft_test_error}
\begin{tabular}{@{}ccccc@{}}
\toprule
\multirow{2}{*}{\textbf{Model}} & \multirow{2}{*}{\textbf{N$_\mathrm{FT{\text -}TS}$}} & \multicolumn{3}{c}{\textbf{Datasets}} \\ \cmidrule(l){3-5} 
 &  & \textbf{$E_{r}$ (M)\cite{test_set_clark,test_set_landers,test_set_li,test_set_peng_role,test_set_peng_trends_2022,test_set_saini_electronic_2022,test_set_snider,test_set_tang_modeling,test_set_tangfrom,test_set_tetteh,test_set_wang,test_set_yang}} & \textbf{$E_{r}$ (MO) \cite{oxide_test_set_flores,oxide_test_set_lee,oxide_test_set_shilong,oxide_test_set_strickler}} & \textbf{$E_{a}$ (M) \cite{oxide_test_set_flores,oxide_test_set_lee,oxide_test_set_shilong,oxide_test_set_strickler}} \\ \midrule
MACE-MH-1 (OMAT) & - & 0.478 & 0.322 & 0.220 \\
FT-BMA & 39,983 & 0.238 & 0.334 & 0.145 \\
FT-MD(M) & 4,976 & 0.376 & 0.428 & 0.149 \\
FT-MD(MO) & 2,692 & 0.334 & 0.306 & 0.184 \\ 
FT-CE & 2,208 & 0.421 & 0.305 & 0.139 \\
FT-BMA+MD(M) & 44,959 & 0.245 & 0.348 & 0.152 \\
FT-All & 49,860 & 0.247 & 0.277 & 0.144 \\ \bottomrule
\end{tabular}
\end{table}




\begin{figure}
\centering
  \centering
  \includegraphics[width=0.5\linewidth]{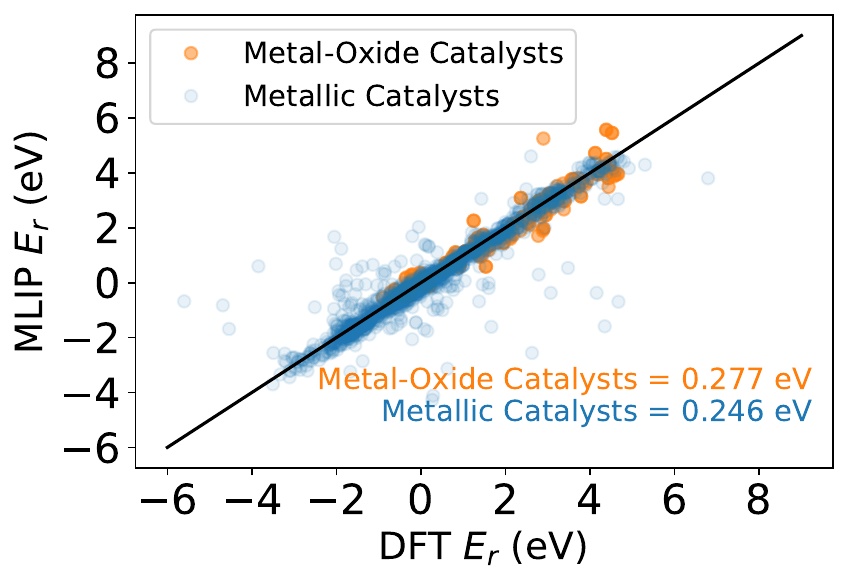}
  \label{fig:parity_test_sub2}
\caption{Parity plot for \reacenergy{} using MLIP FT-All vs DFT for metal-oxide and metallic catalysts test datasets with MAEs shown in annotations.}
\label{fig:parity_test}
\end{figure}

\subsection{Applications}\label{sec:applications}
Now that we have established how training sets influence overall model performance across published studies, we next contrast the performance of different model versions for targeted applications. The systems and structures studied here were a subset of the full test sets discussed in the previous section. These results aim to provide concrete case studies of how a researcher might expect different fine-tuning strategies will impact their MLIP's performance. 

We first focus on the \ce{CO2RR} on metallic catalysts from the two previously published studies \cite{test_set_tangfrom,test_set_peng_trends_2022} and the OER on iridium \cite{oxide_test_set_flores}, as each of these reactions is highly important to electrocatalytic research. We then screen over a large chemical space to test the accuracy of our MLIPs for predicting adsorption energies on binary alloys. 

\subsubsection{\ce{CO2} reduction on metallic catalysts} \label{sec:co2r}
Using \ce{CO2RR} to produce carbon-based products and fuels is one of the great challenges in electrocatalysis. Here, we focus specifically on reactions that lead to \ce{C1} and \ce{C2} products. Peng, et al. \cite{test_set_peng_trends_2022} investigated the selectivity of \ce{CO2RR} toward various \ce{C2} products, including acetylene (\ce{C2H2}), ethane (\ce{C2H6}), ethylene (\ce{C2H4}), and ethanol (\ce{CH3CH2OH}), among others. The protonation of \ce{CCO}* to form \ce{CHCO}* is identified as the only thermodynamically favorable protonation pathway for \ce{CCO}*. The reduction of \ce{CHCO}* can proceed through three distinct intermediates — \ce{CH2CO}*, \ce{OCHCH}*, and \ce{CHCOH}* — with each pathway giving rise to unique or shared byproducts. 

\begin{figure}
\centering
  \centering
  \includegraphics[width=\linewidth]{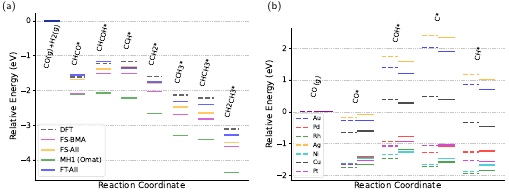}
  \label{fig:co2r_sub2}
\caption{Two reported pathways for the \ce{CO2RR} to (a) \ce{C2} products on Cu(001) \cite{test_set_peng_trends_2022} using the , MACE-MH-1 (OMAT), FS-BMA, FS-All, and FT-All models and (b) \ce{C1} products on 7 different FCC (100) catalysts \cite{test_set_tangfrom} using the FT-All model. The dashed lines are reported DFT values and solid lines are FT-All MLIP values. All pathway energies are reported with respect to gas phase \ce{CO} and \ce{H2} and a clean catalyst surface.}
\label{fig:co2r}
\end{figure}

We validated the MLIPs trained in this work by using them to reconstruct the \ce{CHCOH}* pathway, including all intermediate steps leading to \ce{CH2CH3}*, the final surface-bound species before forming gas-phase ethane on the Cu(001) surface. Fig. \ref{fig:co2r}a contrasts the performance of the MACE-MH-1 (OMAT), FS-BMA, FS-All, and FT-All models for each intermediate step against the previously reported DFT reaction energies. MACE-MH-1 (OMAT) performs the worst with an MAE of 1.011 eV. The FS-BMA model exhibits an error of of up to 0.5 eV for each elementary step (overall MAE of 0.455 eV), but is still capable of correctly ranking the energies of the elementary steps. The error is substantially reduced to 0.251 eV by the FS-All model, which has MD configurations in the training set. However the FT-All model, which was fine-tuned to the same dataset that FS-All was trained to, is the best performing model with an MAE of 0.141 eV across these 7 reactions, and similarly achieve an MAE of 0.143 eV across the entire 101 reactions reported across other catalysts in this study.

Due to the accuracy of the FT-All model, we additionally applied it to predict \ce{CO} protonation to \ce{C1} products via a \ce{COH}* intermediate. Fig. \ref{fig:co2r}b summarizes its performance across all of the studied transition metal catalysts. The FT-All model (solid lines) predicts the energy of each step with an MAE of 0.124 eV with respect to the DFT values (dashed lines) and correctly captures \ce{Ag} as the catalyst with highest \reacbarrier{}, followed by \ce{Au} and \ce{Cu}. We also note that even in cases where the energy difference between catalysts for a state are less than 0.05 eV, e.g. for the C* step on the \ce{Ni} and \ce{Rh} surfaces, the FT-All model can still correctly order the metals, suggesting its can be useful even for energetically similar reaction screening.


To test model performance for finding \ce{CO2RR} transition states, we performed NEB MLIP calculations using MLIP-optimized endpoint structures for \ce{CO2} reduction to \ce{C2} and \ce{C3} products, and compared our MLIP NEB transition states to the previous predicted DFT transition states. \cite{test_set_peng_role,tang_exp}
Tang, et al.\cite{tang_exp} considered CO coupling with \ce{C2} intermediates, including \ce{OCCO}* and \ce{CCO}*, \ce{HCCH},  \ce{CH3CHO}*, \ce{CH2CHO}*, \ce{C2H4}*, etc., on Cu (001) and (511) to form \ce{C3} products. Their work provides 23 DFT NEB reactions created using similar DFT settings as those used to train our MLIPs, and thus are an excellent evaluation set for our models. All FT models, as was already discussed in the broader test set in Table \ref{table:ft_test_error}, have almost the same performance as the FT-All model. For the reactions in Ref. \cite{tang_exp}, even the FT-MD(MO) model, with a fine-tuning dataset of just 2,692 structures on metal-oxide catalysts, has a lower MAE than FS-All model. Therefore, while a from-scratch MLIP needs training set configurations that include bond-breaking events to accurately predict reaction pathways, FT models do not need state-of-the-art sampling techniques. This can be explained by the fact that MACE-MH-1 have a diverse enough training set that fine-tuning to only a small subset of configurations (with or without bond-breaking events or high energy states) can result in robust performance in finding transition states.





\begin{figure}[H]
\centering
\includegraphics[width=1.0\linewidth]{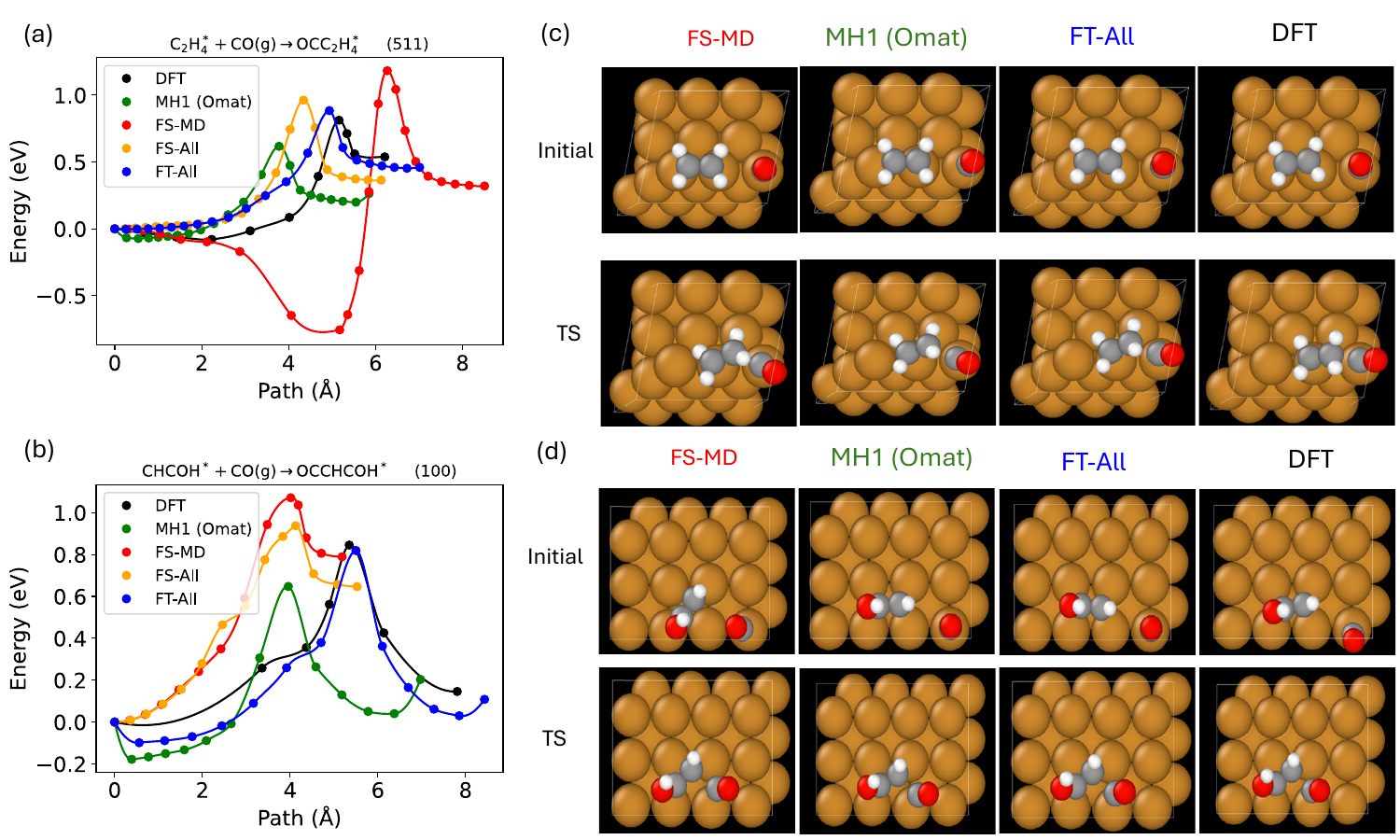}
\caption{The NEB calculated reaction pathways for (a)  \ce{C2H4}* + \ce{CO} $\rightarrow$ \ce{OCC2H4}* on Cu(511) and (b) \ce{CHCOH}* + \ce{CO} $\rightarrow$ \ce{OCCHCOH}* on Cu(100) surfaces, using DFT, MACE-MH-1 (OMAT), FS-MD, FS-All, and FT-All MLIP models. The initial and transition state structures for each model are depicted in (c) and (d).}
\label{fig:neb_fig}
\end{figure}

To further investigate the small differences in MAE of \reacbarrier{}, two example reaction pathways: \ce{C2H4}* + \ce{CO} $\rightarrow$ \ce{OCC2H4}* on Cu(511) and \ce{CHCHO}* + \ce{CO} $\rightarrow$ \ce{OCCHCHO}* on Cu(100) were chosen to contrast DFT and 4 versions of our trained models (Fig. \ref{fig:neb_fig}). For each pathway, the NEB calculation is performed independently with the assigned calculator. One of the challenges in comparing MLIP and DFT performance is in how they can relax the endpoints into different local minima, which thus impacts the subsequent transition state search. In addition, the minimum energy pathway from one MLIP can differ from another, even with almost same endpoints, and one may get different transition state structures. To show the structural similarities and differences, we depicted the top view of both reactions at initial and transition states from DFT and the FS-MD, MACE-MH-1 (OMAT) models in Fig. \ref{fig:neb_fig}c-d. For the \ce{C2H4}* + \ce{CO} $\rightarrow$ \ce{OCC2H4}* reaction, except for slight shift in adsorption site \ce{C2H4} or the orientation of \ce{CO}, the initial state of both \ce{C2H4} and \ce{CO} are very similar across all four MLIPs. However, at the transition state, the FS-MD model differ significancy from other MLIPs, with \ce{C2H4} rotated in-plane such that one hydrogen is closer to C end of \ce{CO}. For the other reaction, \ce{CHCHO}* + \ce{CO} $\rightarrow$ \ce{OCCHCHO}*, at the initial structure, the FS-MD model optimized the CHCOH to a rotated state, with more than 40 degrees azimuthal rotation compared to all other models. On the other hand, the transition states across all 4 models are almost identical. In another words, the 0.2 eV higher error in predicting the barrier by the FS-MD model arises because it has predicted a different initial state as compared to DFT, not the transition state. This example highlights that only using MAEs to compare models may not provide a full description of model performance for barrier predictions because they require models to have simultaneous accuracy for both the energies and structures of initial and transition states on the PES. To better illustrate this, we applied FT-All to the final optimized images of each NEB calculations (SI Section S6), and observed a difference of 0.3 eV in barrier predictions, indicating a high sensitivity of \reacbarrier{} to the structures at the initial and transitions states. 


\subsubsection{Oxygen evolution reaction on iridium oxides} \label{sec:oer}
Next, we evaluate the performance of our models to predict OER elementary steps on different metal-oxide surfaces. In particular, iridium oxide polymorphs are important class of metal-oxides that have widely been studied for the acidic OER. Flores, et al. \cite{oxide_test_set_flores} calculated the OER energetics for a total of 96 reactions on rutile (R), $\alpha$, and $\beta$ phases of several different surface terminations of \ce{IrO2} and \ce{IrO3} under applied potential. 


We first compare the performance of the MACE-MH-1 (OMAT), FS-All, FT-MD(MO), and FT-All models to DFT for the OER on \ce{$\alpha$-IrO3}(110) fully covered with OH. The FS-All model, trained to 2,692 metal-oxide configurations in addition to 47,168 metallic surfaces, has an error of 0.21 eV for O and OH adsorption. However, the error is significantly higher for OOH* adsorption at 1.6 eV, which is attributed to the absence of OOH* species in the training set. For the 96 reactions reported in Ref. \cite{oxide_test_set_flores}, the FS-All model has an MAE of 0.438 eV. Using the MACE-MH-1 model (no fine-tuning) reduces the error for OOH* to less than 0.5 eV, and has an overall MAE of 0.384 eV, which is a 0.05 eV lower than FS-All model. We attribute this difference  to the very large dataset of solid-state metal-oxides in the OMAT dataset. Similarly, UMA (OMAT) also performs well because of the OMAT dataset is in the training set (see Table \ref{table:test_error}). However, fine-tuned MLIPs, as was discussed in Sec. \ref{sec:fine-tune}, have a noticeably improved performance compared to these foundation models. Here we compare the FT-MD(MO) model, which is trained to only 2,692 metal-oxide configurations, and the FT-All model, which is fine-tuned to our entire dataset including metallic and metal-oxide catalysts. For this application, both models exhibit excellent performance for O and OH adsorption, with errors of less than 0.1 eV, but still struggle predicting the OOH* step, indicating that the OOH* step is highly underrepresented in the overall dataset of large foundation models. The total MAEs for the 96 reaction for FT-MD(MO) and FT-All models are 0.313 and 0.278 eV, respectively, both showing better performance than the from-scratch models.

\begin{figure}
  \centering
  \includegraphics[width=\linewidth]{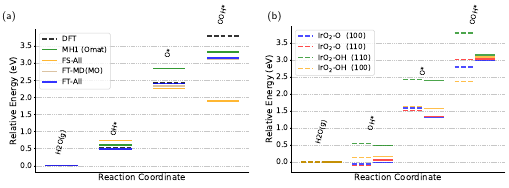}
\caption{(a) Comparison of the MACE-MH-1 (OMAT), FS-All, FT-MD(MO), and FT-All MLIPs (solid lines) for OER pathway predictions on OH covered $\alpha$-\ce{IrO3} with the DFT reported values (dashed lines). (b) Comparison of the FT-All model performance (solid lines) OER pathway prediction on O and OH covered rutile-\ce{IrO2} (100) and (110) surface terminations with DFT values (dashed lines).}
\label{fig:OER}
\end{figure}

Next, we predicted OER intermediate energies using the FT-All model for the OH and O covered surfaces of \ce{IrO2} (100) and (110). Fig. \ref{fig:OER}b shows that, except for two outliers for the OOH* step in OER, the MLIP predictions are within 0.15 eV of the reported DFT values, and in most cases, predict the same energy ordering across different terminations as DFT, similar to the behavior observed in Section \ref{sec:co2r}. For example, both DFT and the FT-All model predict that additional \ce{OH}* formation on OH-covered \ce{IrO2} (110) and (100) surfaces is less favorable than on the O-covered surfaces.



\subsubsection{Screening \screentotalsamples{} adsorption energies on binary alloys}

A key advantage of MLIPs over DFT is that they enable rapid screening a large chemical space while still retaining good accuracy. Given the reliability of the FT-All MLIP discussed in previous sections, here we applied it to predict adsorption energies on binary metal surfaces selected from 8 common FCC metal alloys in 3 different mixing ratios and 6 possible surface terminations. Specifically, we screened over alloys of Ni, Cu, Au, Ag, Ir, Pd, Pt, and Rh with 100/0\%, 25/75\%, and 50/50\% mixing ratios and the (001), (111), (211), (311), (511), and (532) facets. Then we placed 33 different adsorbates on all unique binding sites on the surfaces in an approach described further in SI section S8. This procedure produced \screentotalsamples{} total initialized structures, of which \screenconvergedsamples{} converged. The FT-All model was used for all parts of the adsorption energy workflow, namely bulk structure optimization, surface cleaving, and adsorbate relaxation. 

\begin{figure}
\centering
\includegraphics[width=\linewidth]{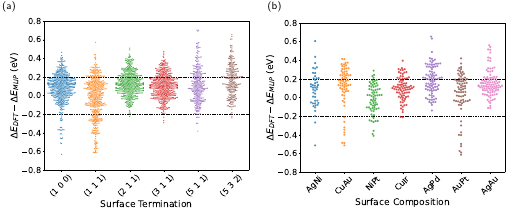}
\caption{ Bee-swarm plots for randomly selected test set adsorption energy predictions grouped by (a) surface termination and (b) surface composition in decreasing order of lattice mismatch. The dashed red lines indicates the 0.2 eV threshold for the acceptable accuracy for the adsorption energy error.}
\label{fig:screen_ads}
\end{figure}

To evaluate model performance, we selected \screentestcases{} structures that are almost uniformly distributed across surface terminations, adsorbate used, and surface compositions, and performed DFT calculations using the same settings as described above. For these \screentestcases{} configurations, the FT-All model adsorption energy MAE is \screensubsampleError{} eV, which is similar to the errors reported in earlier sections and in Table \ref{table:test_error}. Next, we further analyzed this error by calculating it on subsets of this test set grouped by surface composition, surface termination, and the types of the adsorbates. Fig. \ref{fig:screen_ads}a shows that the adsorption energy error only somewhat increases for the higher Miller index surfaces. For the (311) surface termination, 77\% of configurations are within the target threshold of $\pm 0.2$ eV. The accuracy is lowest for the (532) facet, the highest surface termination we considered, with only 58\% of predictions within the error threshold. Nevertheless for all other surface terminations, 65\% to 75\% of the tested configurations lie within with the $\pm 0.2$ eV tolerance.

As an another test of model performance, seven surface compositions were selected out of the 56 total surface compositions to compare their adsorption energy errors. Here, surface compositions were selected to span element pairs with the highest atomic radii mismatch (AgNi) to those with nearly identical atomic radii (AgAu). This tests if error might depend on the degree to which a surface can undergo relaxations/reconstruction due to radii differences. For AgNi and CuAu, the performance is mediocre with only 68\% and 46\% within $\pm 0.2$ eV of the DFT values, respectively. On the other hand, the model performance is much better for CuIr, with 73\% within the threshold. Overall, we find that the percentage of acceptable cases does not strongly correlate with atomic radii mismatch, presumably because other factors, such as the adsorbate types, surface terminations, and mixing ratios of alloys, also affect the error of the model. Nevertheless, the model performs well across majority of the cases, and if a higher accuracy is needed for a targeted adsorption energy, additional configurations from relaxation trajectories can improve the model performance.


\section{Conclusion}\label{sec13}
This study provides an evaluation of MACE models trained from-scratch and MLIPs fine-tuned to the MACE-MH-1 foundation model for prediction of catalytic processes. We used different sets of training data, including relaxation trajectories and MD or CE sampling techniques, to show the dependence of MLIP performance in predicting in \reacenergy{} and \reacbarrier{} for a set of metallic and metal-oxide catalyst reactions. In particular, for from-scratch models, perturbed configurations from MD or CE simulations plays a key role in reducing the error. However, they become less important when fine-tuning from MACE-MH-1, a pretrained model that already exhibits decent performance for catalysis. Our results show that our MLIPs fine-tuned using metallic catalyst data with only relaxation trajectories and single atom adsorbates can also be used to predict metal-oxide reactions with reasonable accuracy. For \reacenergy{} of \ce{CHCOH} pathway for the \ce{CO2}RR, the FT-All model achieves an MAE of 0.141 eV, noticeably better than our best from scratch model (FS-All MAE = 0.251 eV). For the performance on \reacbarrier{}, we performed NEB calculations for a class of \ce{CO2} reduction reactions that involve \ce{CO} coupling to a range of intermediate \ce{C2} products. Our results indicate all our MLIPs fine-tuned to a different training set each have a MAE of between 0.14 to 0.15 eV, better than our best performing from scratch model (MAE = 0.175 eV). Additionally, we applied our MLIPs to study OER on \ce{IrO} polymorphs as an example of reactions on metal-oxide surfaces. Our results show that FT-All model has an MAE of 0.278 eV, noticeably better the error of from-scratch model (MAE = 0.384 eV). For reaction pathways on either metallic or metallic catalysts, our results shows that one can successfully predict the \reacenergy{} across different facets or catalysts type with almost same ordering in energy as in DFT calculations. Lastly, we applied the FT-All model to screen adsorption energy predictions for underrepresented surface terminations, compositions, and reactions with 81\% of randomly selected subset of reactions having less than $\pm 0.2$ eV error with respect to the DFT values.

The results of this study demonstrates the efficiency and accuracy of fine-tuning to large models, in particular MACE-MH-1, compared to training from-scratch, and shows that with small datasets, even with different chemical environment, one can train a highly generalizable MLIP that can be applied for screening catalytic reactions.

\section{Supplementary information}
The supporting information includes (1) atomic distribution of bim-metallic datasets (2) Performance of contour exploring (3) parity plots for all training set reaction energies (4) parity plots for all test set reaction energies (4) All NEB comparison pathways and (5) List of all adsorbates in the screening task.

\section{Data Availability}

All the DFT calculations and MLIPs trained in this study, both from scratch and FT models are provided in the Zenodo directory 10.5281/zenodo.20043317 

\section{Acknowledgments}

This work was supported by the U.S. Department of Energy, Office of Science, Basic Energy Sciences, under Award No. DE-SC0022247, as well as funding provided to the University of Toronto's Acceleration Consortium from the Canada First Research Excellence Fund (Grant number - CFREF-2022-00042).
Calculations were carried out at the National
Energy Research Scientific Computing Center (NERSC), a U.S. Department of Energy Office of Science User Facility located at Lawrence Berkeley National Laboratory, operated under Contract No. DE-AC02-05CH11231 using NERSC award ERCAP0020105. 

\section*{Declarations}

GC is a partner in Symmetric Group LLP that licenses force-fields commercially and also has equity interest in Angstrom AI.

\appendix
\renewcommand{\thesection}{S\arabic{section}}
\setcounter{section}{0}
\section*{Supporting Information}

\section{Atomic distribution for bi-metallic alloy dataset \cite{bma}}
\begin{figure}[H]
    \centering  
\includegraphics[width=\textwidth]{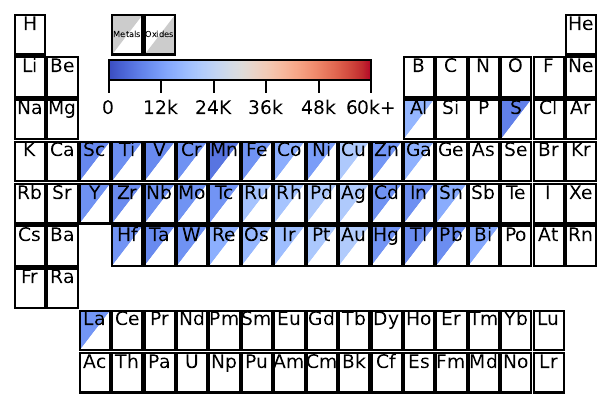}
    \caption{ }
\end{figure}

\clearpage
\section{Performance of Contour Exploring}
\begin{figure}
    \centering  
\includegraphics[width=\textwidth]{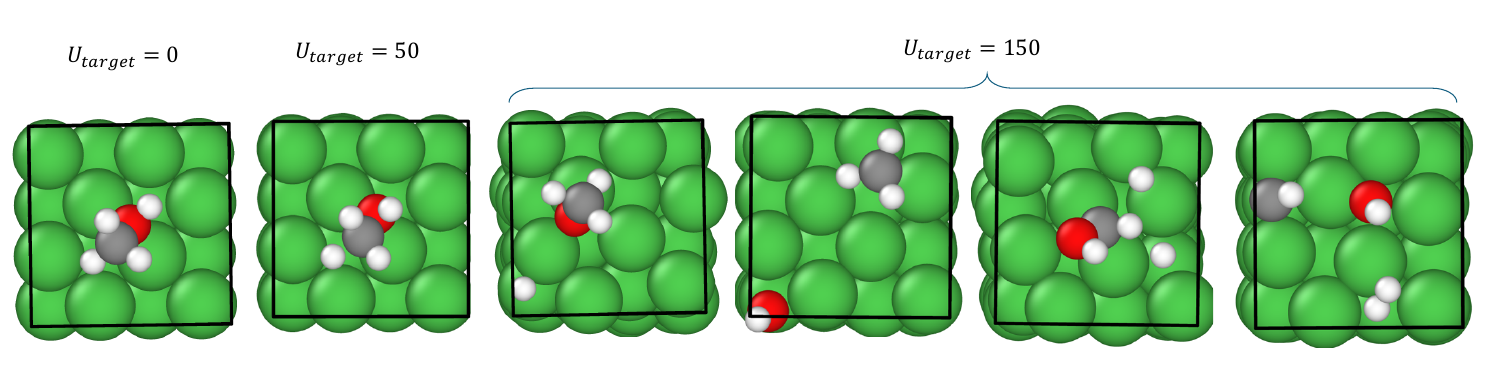}
    \caption{\label{fig:CE}Top-view snapshots of 6 configurations taken at the ground state (left snapshot with $U_{target}$=0) to the 4 independent CE runs with $U_{target}$=150 meV/atom. In all cases, the snapshot was taken at the last step of the trajectory.}
\end{figure}

For example, for arbitrary adsorption of \ce{CH3OH} on Ni (001) surface, we adjusted the $U_{target}$ from 0 to 150 meV/atom, depicted 6 snapshots of their trajectory in Fig.\ref{fig:CE} At $U_{target}=50$ meV/atom, no bond-breaking occurs but still noticeable changes to bond lengths and bond angles occur. At $U_{target}=150$ meV/atom, multiple bond-breaking events happen, and in all these four cases, only the initialization of velocities are different. This suggest that in 4 different runs, we were able to achieve 4 possible bond-breaking events that lead to H, OH, CHOH, CH and gas phase \ce{H2} (\ce{H2} has desorbed from the surface).


\clearpage
\section{Training Set Applications}

In this section, we have plotted the parity between the predicted reaction energy using v5 MLIP and the DFT reaction energies (reported reaction energy) for each dataset used in this study.

\subsection{Selectivity of Synthesis Gas Conversion to C2+ Oxygenates on fcc (111) Transition-Metal Surfaces \cite{schumann}}
\begin{figure}[H]
    \centering  
\includegraphics[width=\textwidth]{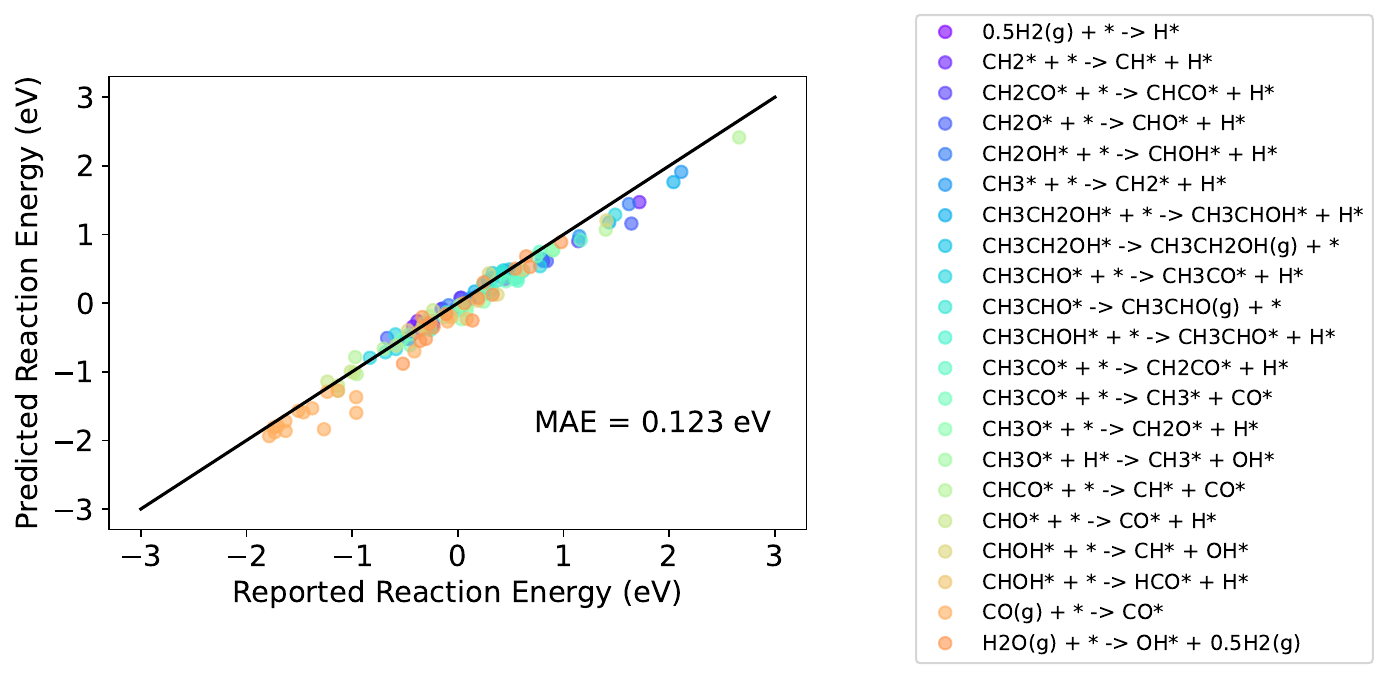}
    \caption{}
\end{figure}

\subsection{Non-oxidative Dehydrogenation of Ethane over Close-packed Metallic Facets \cite{HANSEN2019161}}
\begin{figure}[H]
    \centering  
\includegraphics[width=\textwidth]{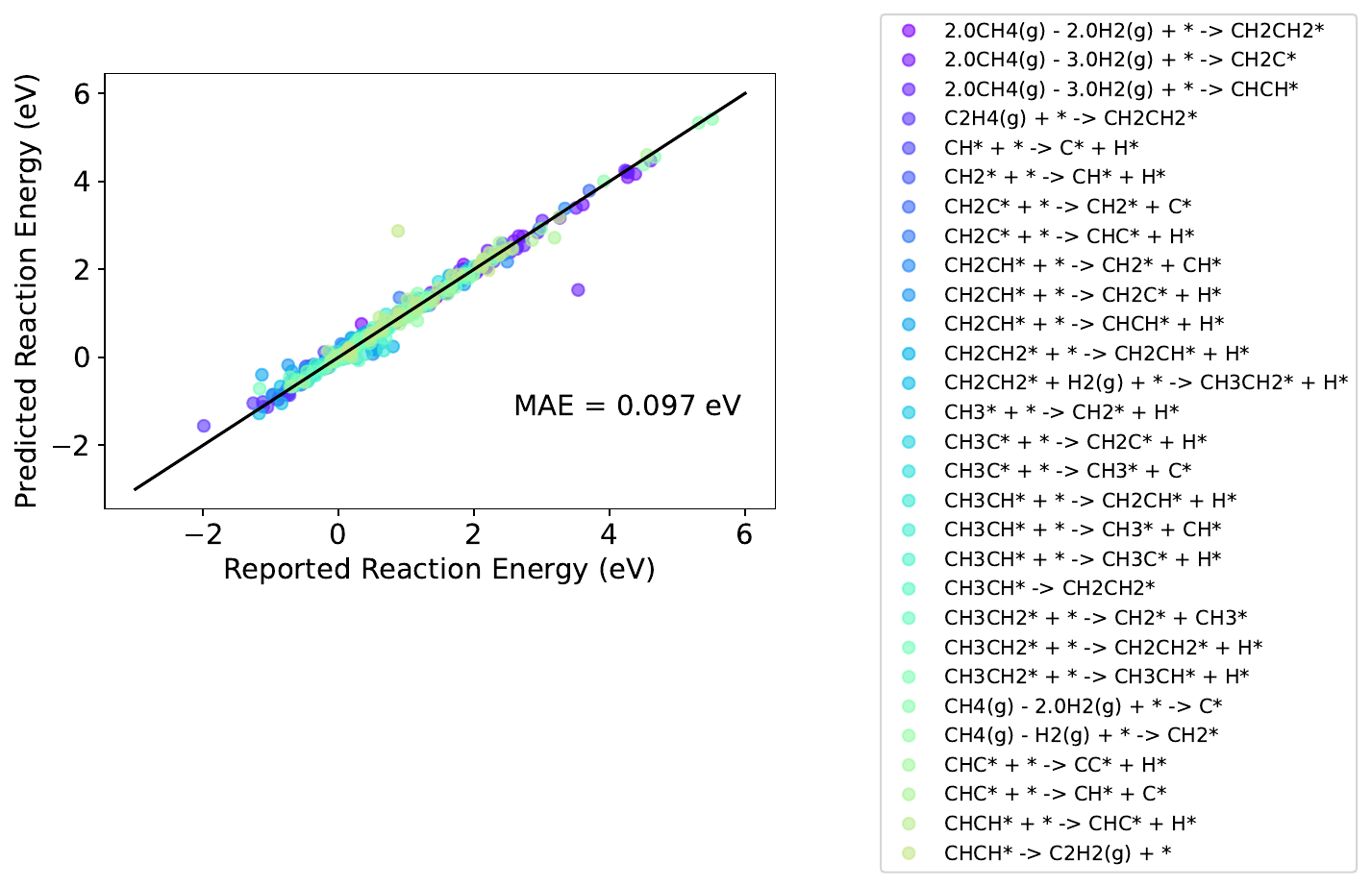}
    \caption{}
\end{figure}

\subsection{Electrochemical Ammonia Synthesis \cite{montoya}}
\begin{figure}[H]
    \centering  
\includegraphics[width=\textwidth]{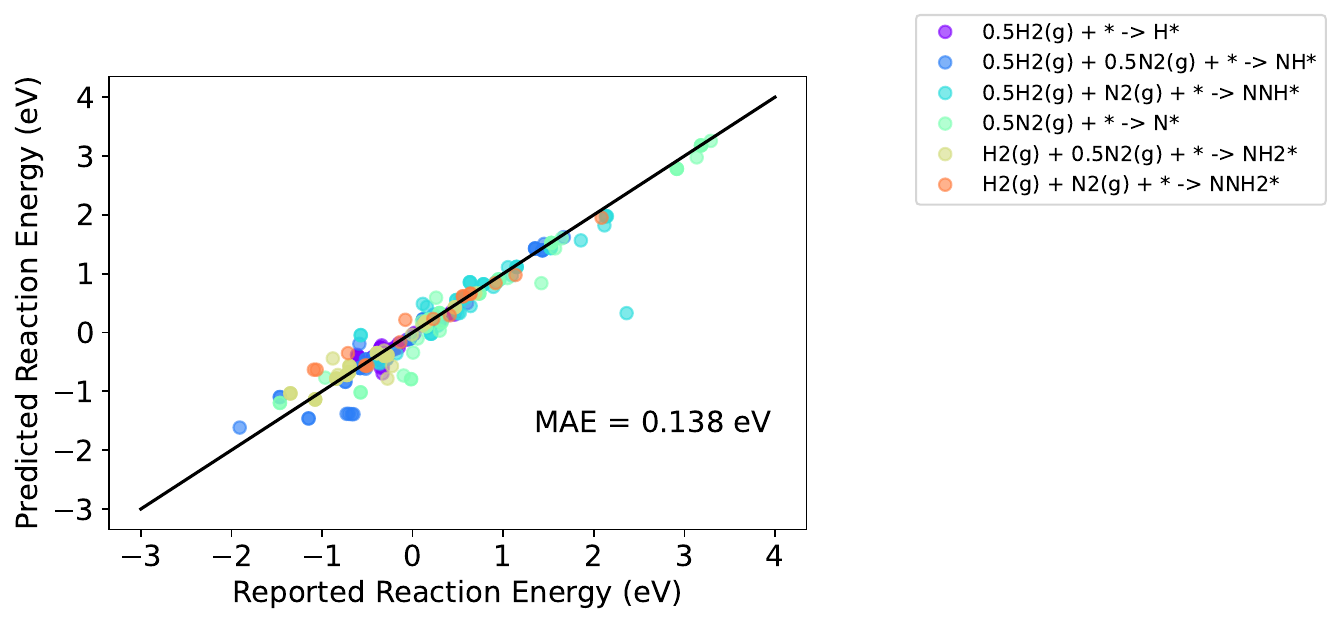}
    \caption{}
\end{figure}

\subsection{Industrial Ammonia Synthesis Rates at Near-ambient Conditions \cite{Wangachieveing}}
\begin{figure}[H]
    \centering  
\includegraphics[width=\textwidth]{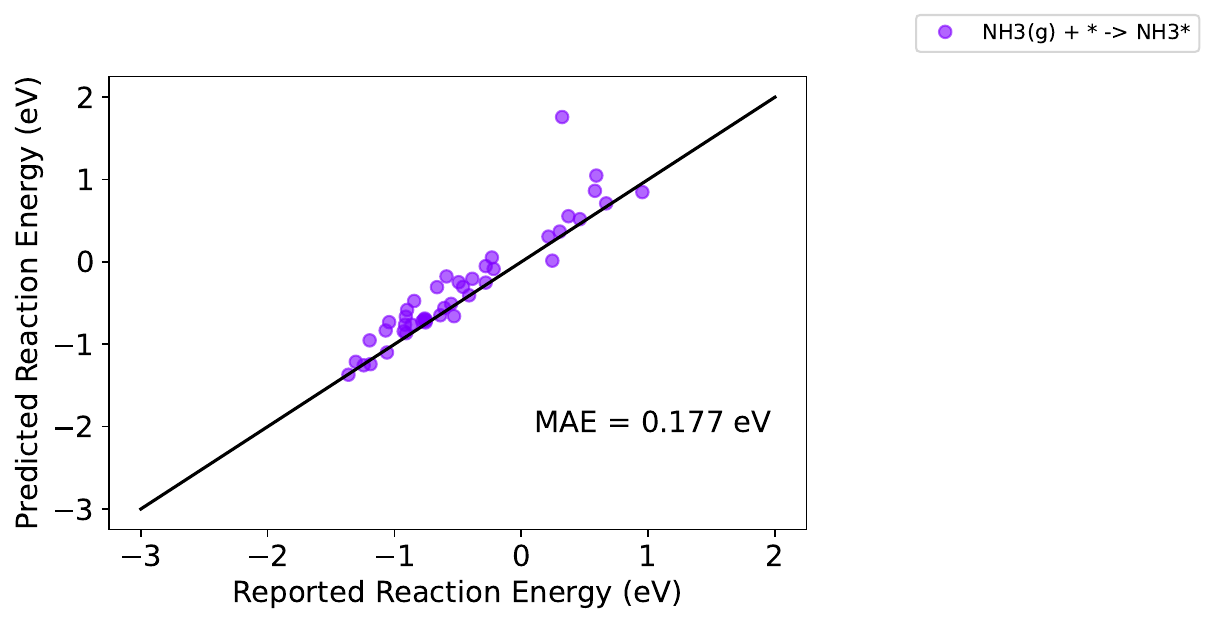}
    \caption{}
\end{figure}

\subsection{Adsorption of O and OH on Transition Metal Oxide Surfaces \cite{comer}}
\begin{figure}[H]
    \centering  
\includegraphics[width=0.5\textwidth]{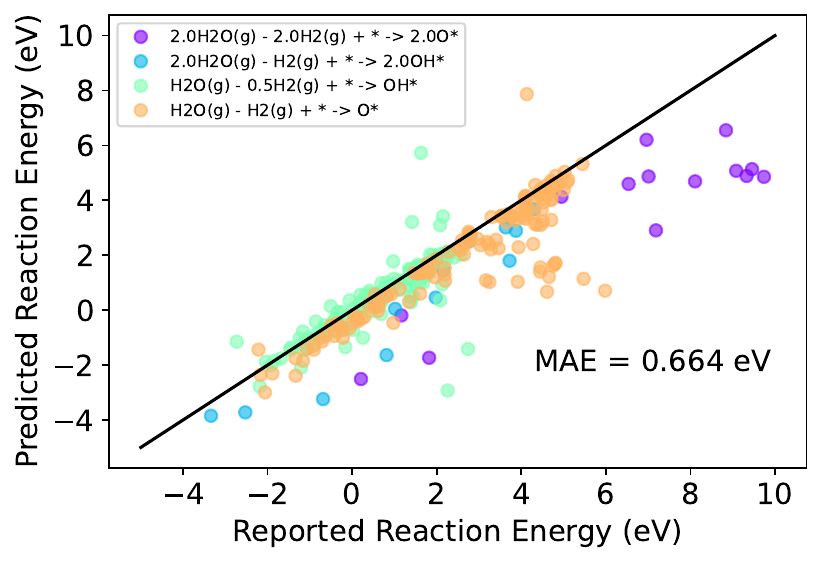}
    \caption{}
\end{figure}

\clearpage
\section{Test Set Applications}
\subsection{\ce{CO2} reduction to other carbon-based products}
\subsubsection{\ce{CO2} to \ce{CO} reduction on Ag surfaces \cite{test_set_clark}} 

\begin{figure}[H]
    \centering  
\includegraphics[width=0.5\textwidth]{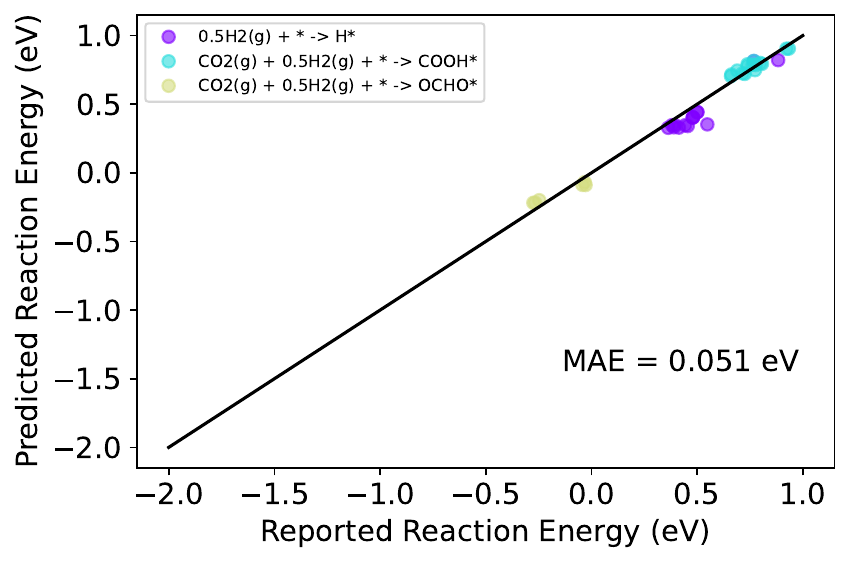}
    \caption{}
\end{figure}

\subsubsection{\ce{CO2} reduction to multi-carbon products on single metal surfaces \cite{test_set_li}} 
\begin{figure}[H]
    \centering  
\includegraphics[width=0.5\textwidth]{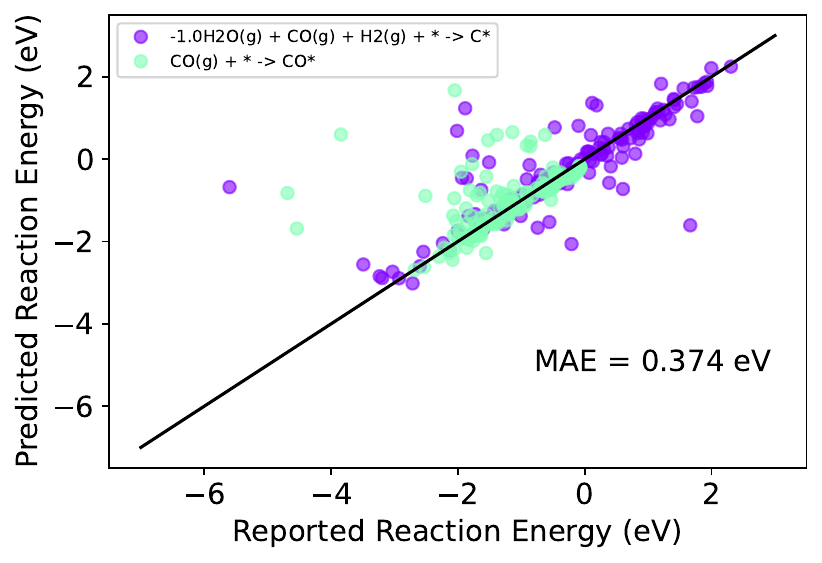}
    \caption{}
\end{figure}

\subsubsection{oxygenate/hydrocarbon selectivity for electrochemical CO(2) reduction to C2 products \cite{test_set_peng_trends_2022}} 
\begin{figure}[H]
    \centering  
\includegraphics[width=\textwidth]{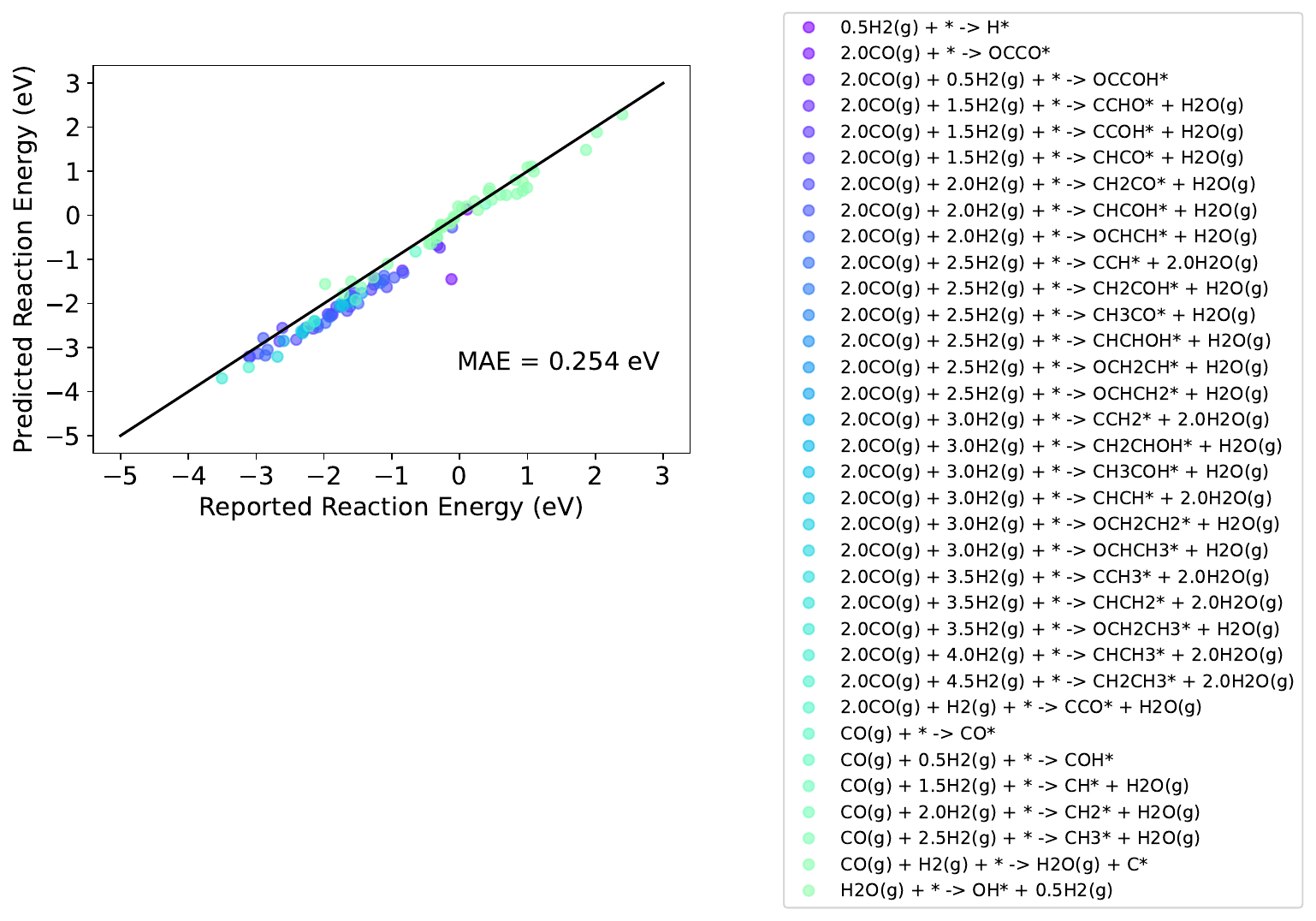}
    \caption{}
\end{figure}

\subsubsection{\ce{CO2} reduction to multi-carbon producst on single metal surfaces \cite{test_set_peng_role}}
\begin{figure}[H]
    \centering  
\includegraphics[width=\textwidth]{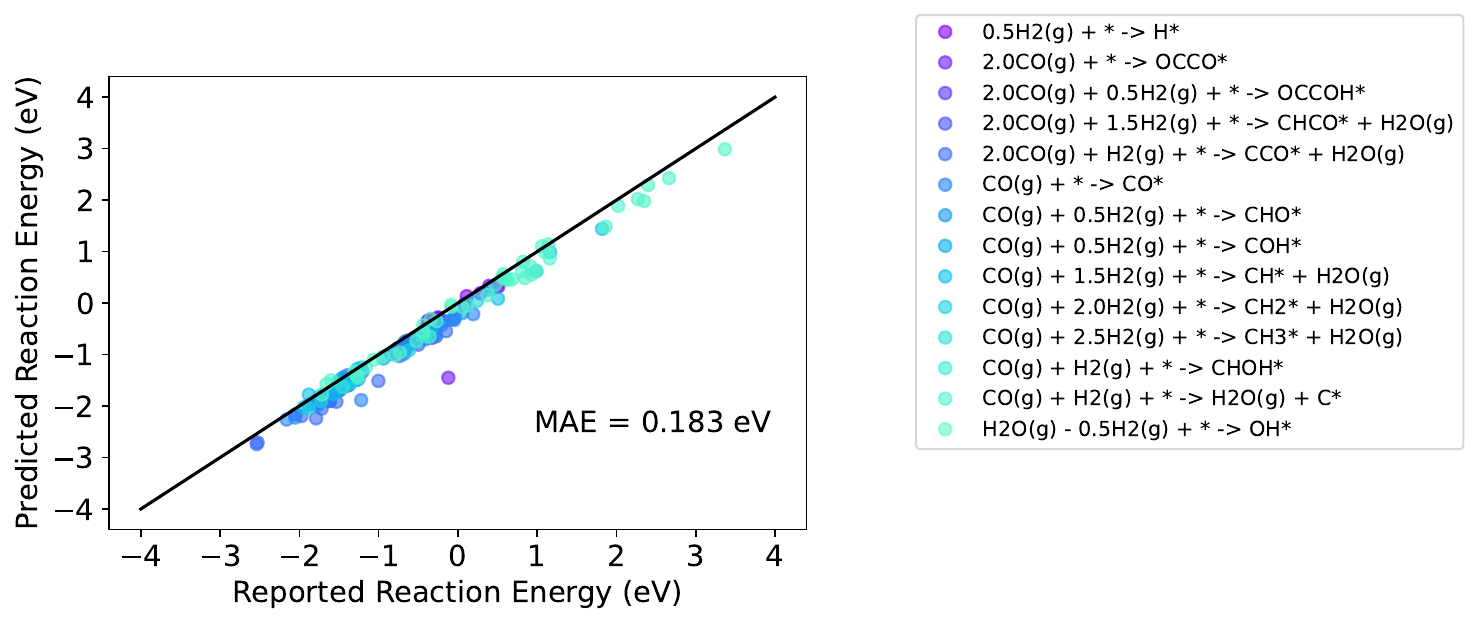}
    \caption{}
\end{figure}

\subsubsection{C1 Selectivity in Electrochemical CO2 Reduction \cite{test_set_tangfrom}}

\begin{figure}[H]
    \centering  
\includegraphics[width=\textwidth]{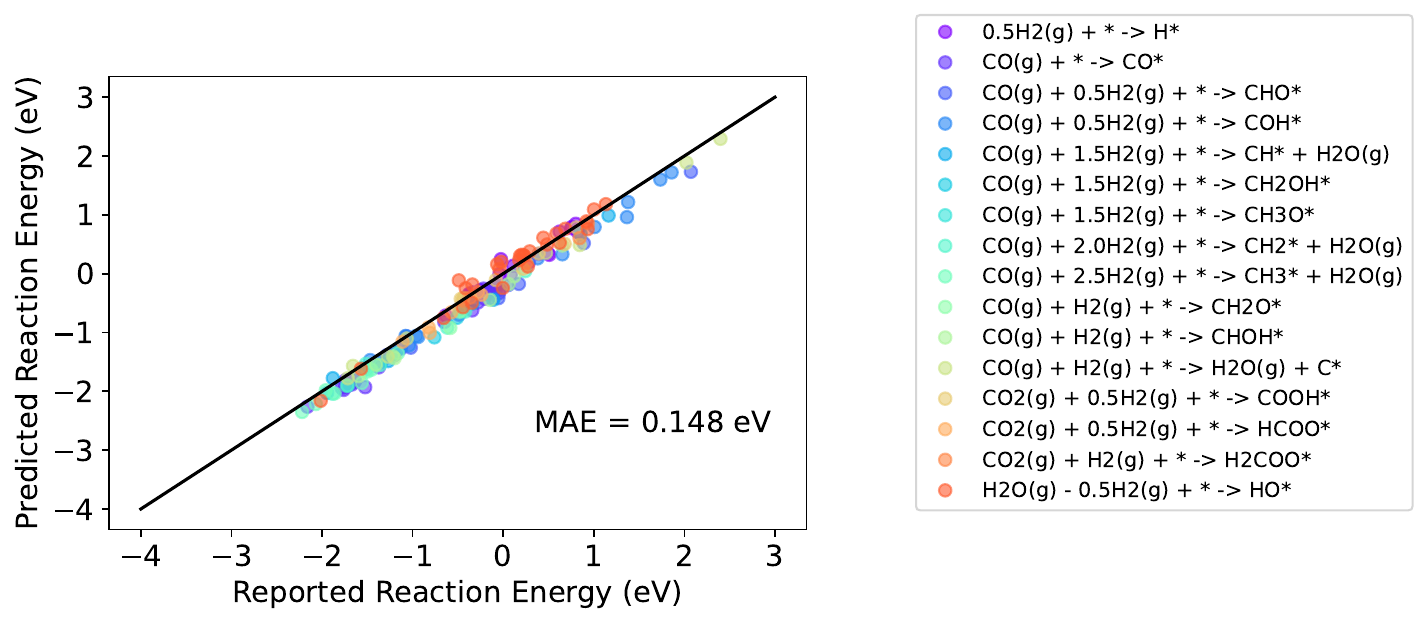}
    \caption{}
\end{figure}

\subsection{Hydrogen Intercalation and Deintercalation in Palladium Electrodes \cite{test_set_landers}} 

\begin{figure}[H]
    \centering  
\includegraphics[width=\textwidth]{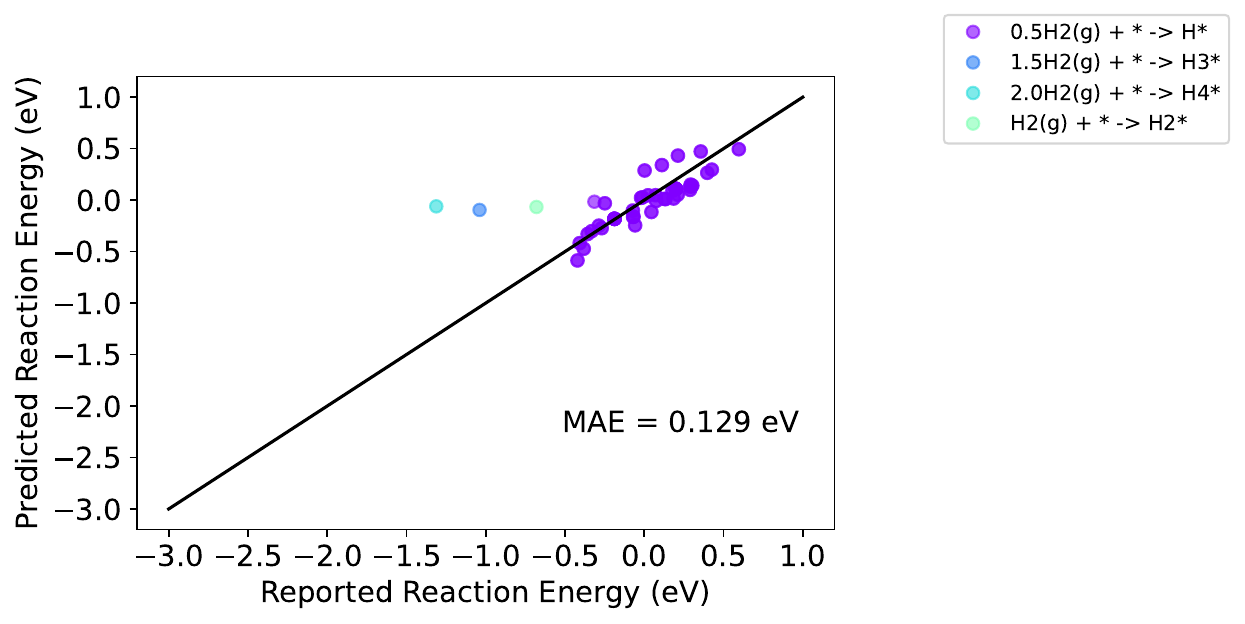}
    \caption{}
\end{figure}

\subsubsection{Adsorption of O, N, CH, and Li on bi- and tri-metallic surface \cite{test_set_saini_electronic_2022}} 
\begin{figure}[H]
    \centering  
\includegraphics[width=0.5\textwidth]{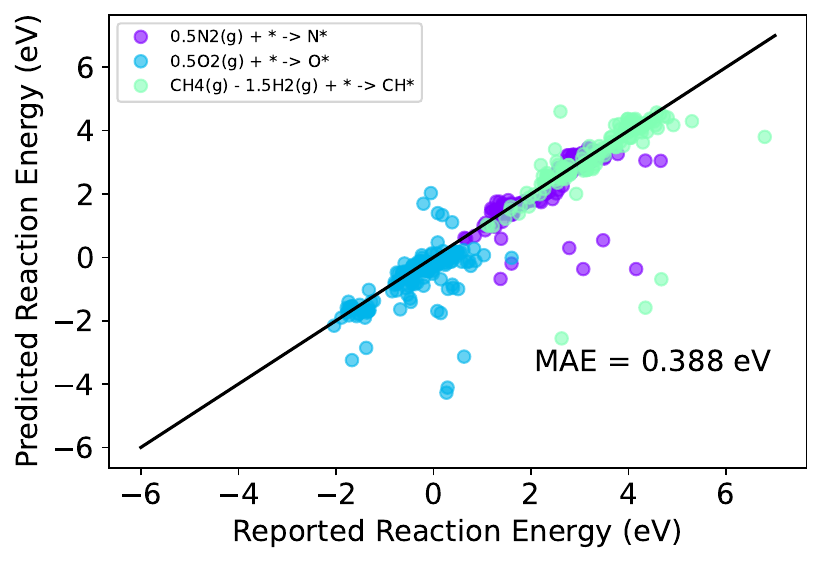}
    \caption{}
\end{figure}
\subsection{\ce{CO2} hydrogenation to methanol \cite{test_set_snider}}

\begin{figure}[H]
    \centering  
\includegraphics[width=\textwidth]{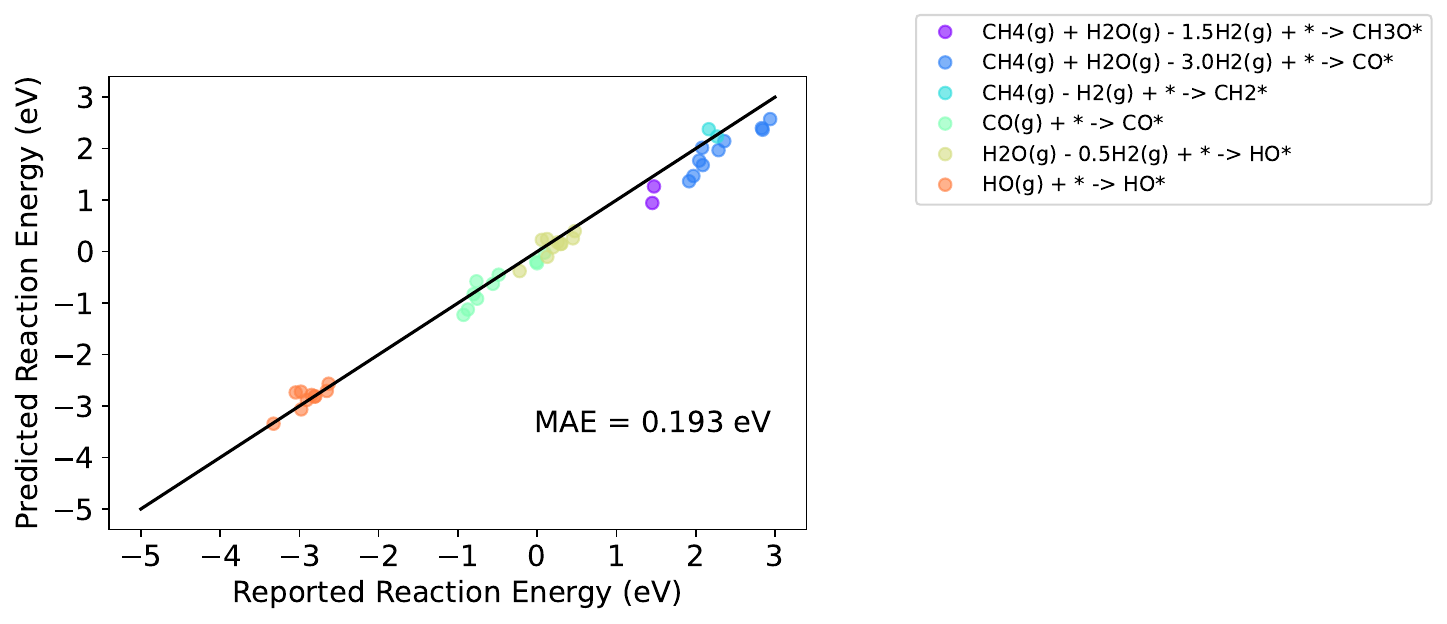}
    \caption{}
\end{figure}

\subsection{Modeling Hydrogen Evolution Reaction Kinetics through Explicit Water–Metal Interfaces \cite{test_set_tang_modeling}}

\begin{figure}[H]
    \centering  
\includegraphics[width=0.5\textwidth]{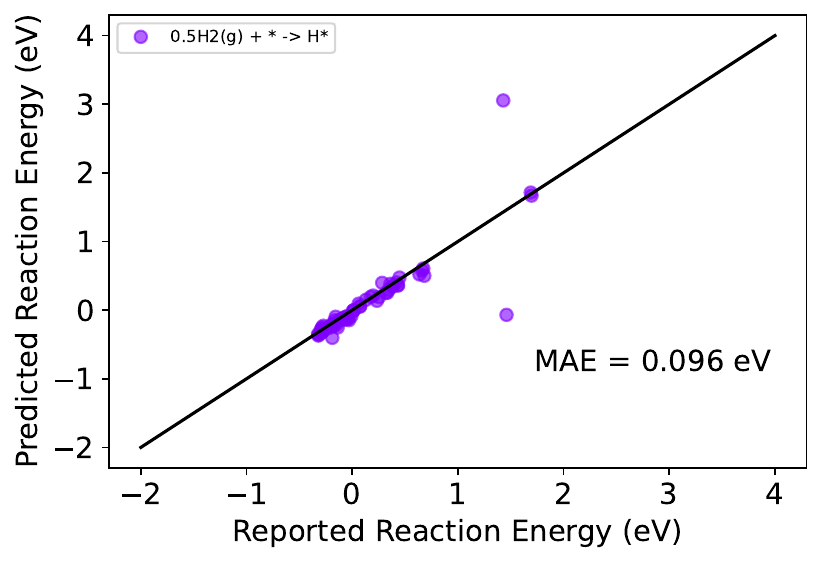}
    \caption{}
\end{figure}

\subsection{Oxygen Reduction and Hydrogen Evolution Activity on Pt \cite{test_set_tetteh}}
\begin{figure}[H]
    \centering  
\includegraphics[width=\textwidth]{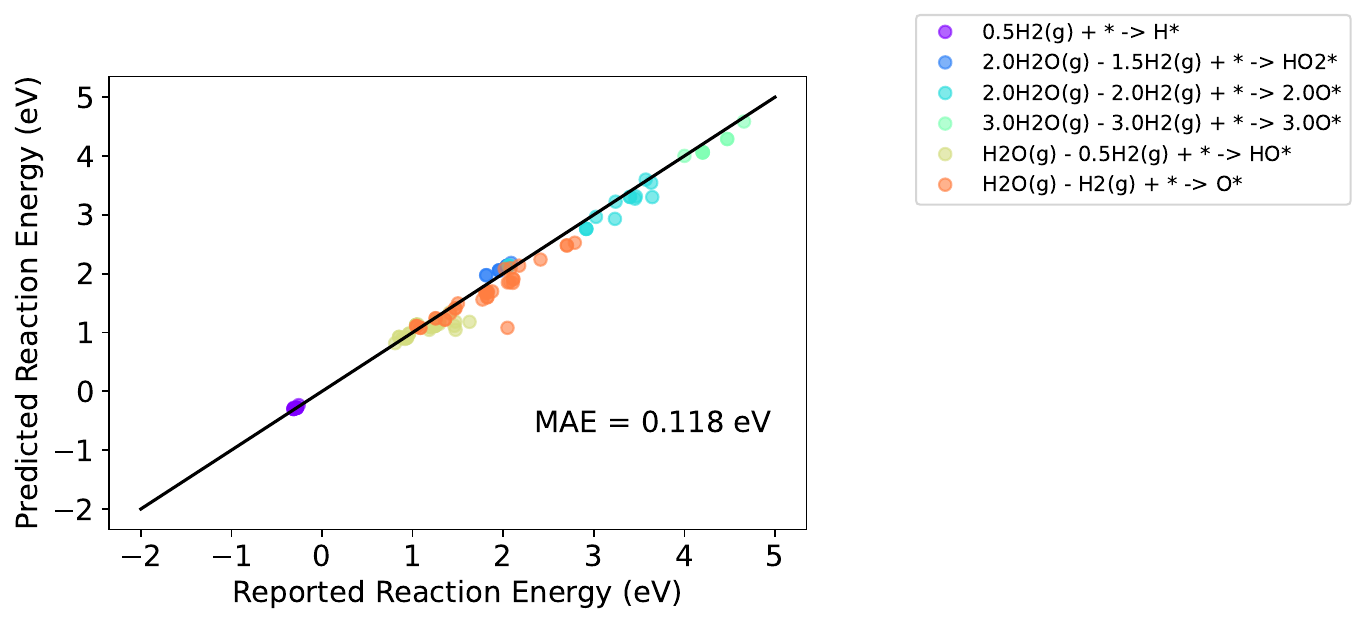}
    \caption{}
\end{figure}

\subsection{Propane Dehydrogenation to Propylene \cite{test_set_wang}}
\begin{figure}[H]
    \centering  
\includegraphics[width=\textwidth]{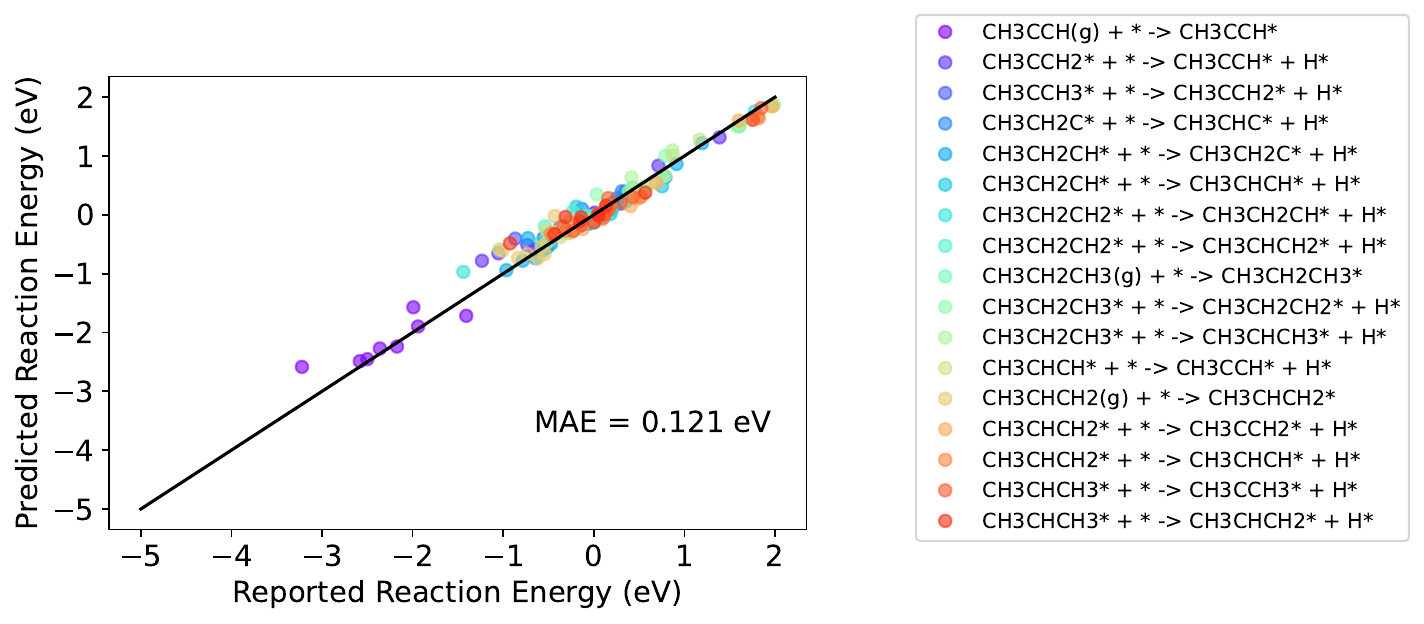}
    \caption{}
\end{figure}

\subsection{Rhodium Catalysts for C2+ Oxygenate Production \cite{test_set_yang}}
\begin{figure}[H]
    \centering  
\includegraphics[width=\textwidth]{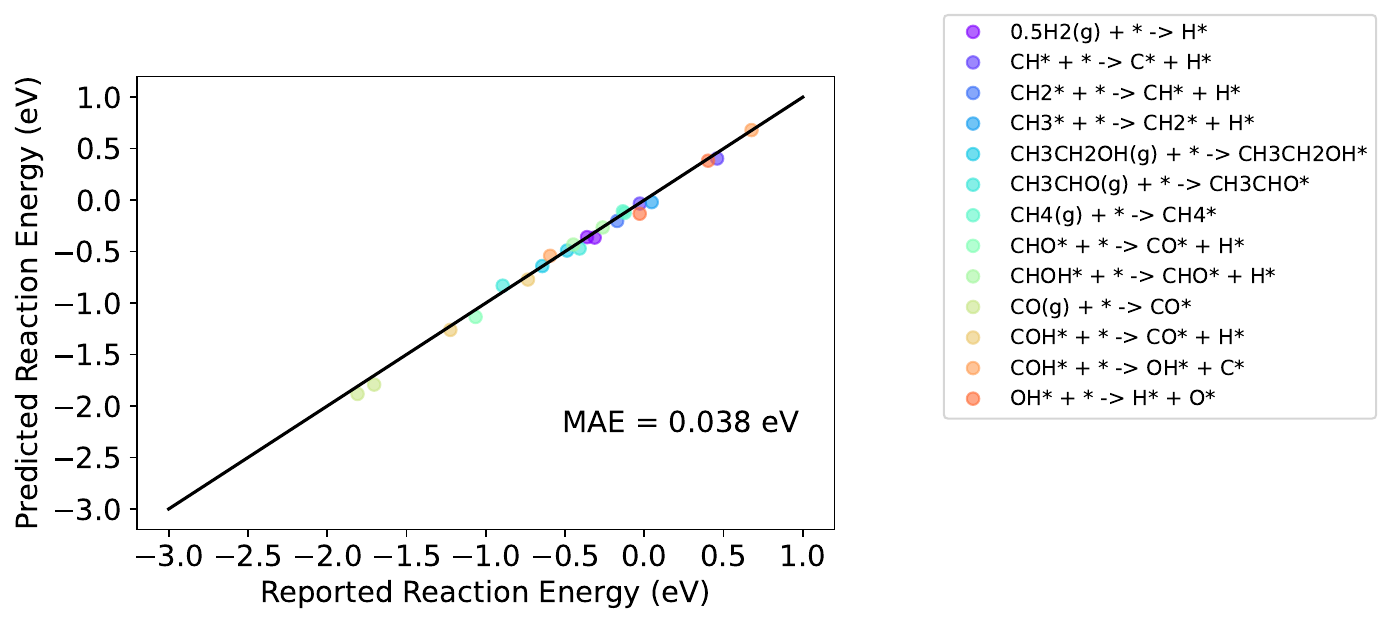}
    \caption{}
\end{figure}



\clearpage
\onecolumn

\section{NEB calculations}
Here we compare the NEB reaction pathways for each MLIP against the DFT values.

\subsection{\ce{CO2} reduction to \ce{C2} products \cite{test_set_peng_role}}

\begin{center}
\begin{minipage}{0.95\textwidth}
\centering

\begin{minipage}[b]{0.48\textwidth}
    \centering
    \includegraphics[width=\linewidth,height=0.19\textheight,keepaspectratio]{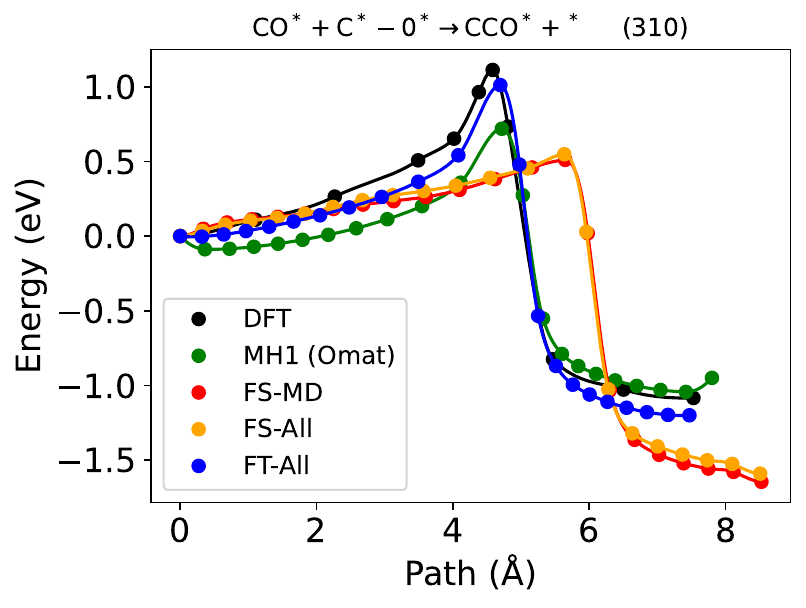}
\end{minipage}\hfill
\begin{minipage}[b]{0.48\textwidth}
    \centering
    \includegraphics[width=\linewidth,height=0.19\textheight,keepaspectratio]{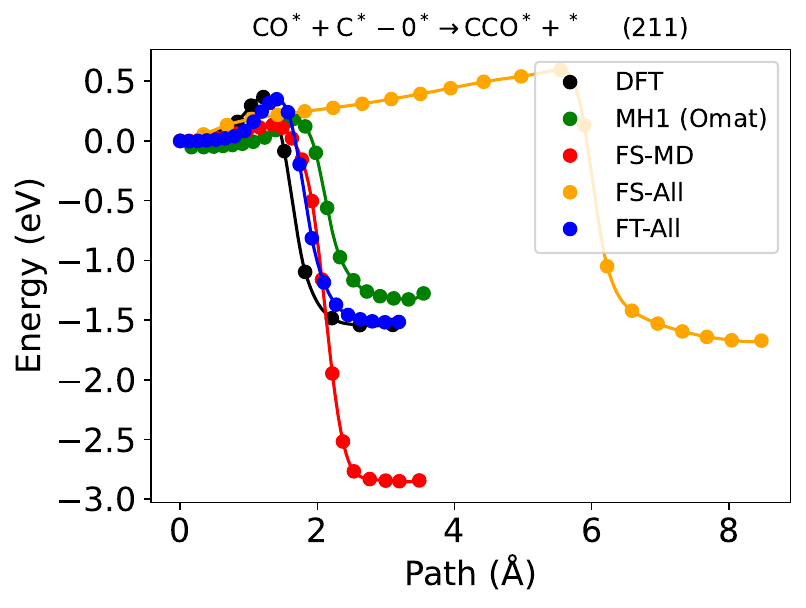}
\end{minipage}

\vspace{4pt}

\begin{minipage}[b]{0.48\textwidth}
    \centering
    \includegraphics[width=\linewidth,height=0.19\textheight,keepaspectratio]{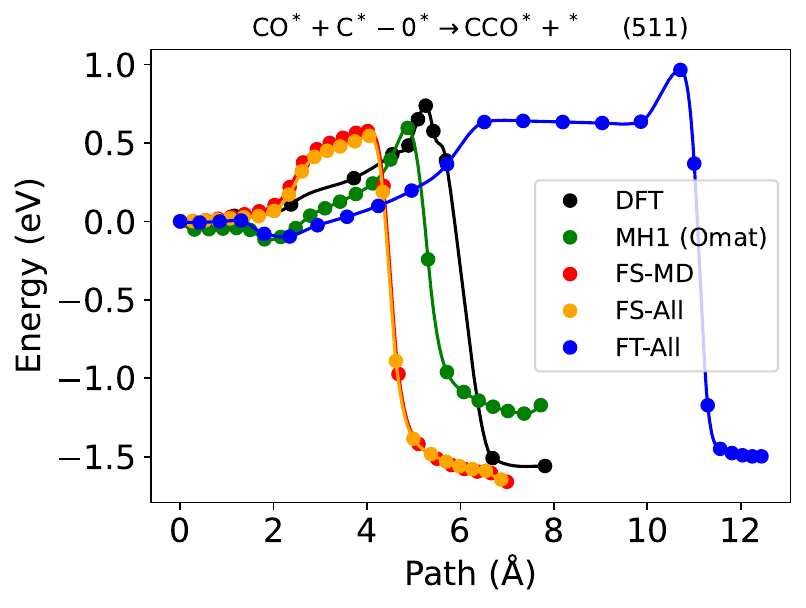}
\end{minipage}\hfill
\begin{minipage}[b]{0.48\textwidth}
    \centering
    \includegraphics[width=\linewidth,height=0.19\textheight,keepaspectratio]{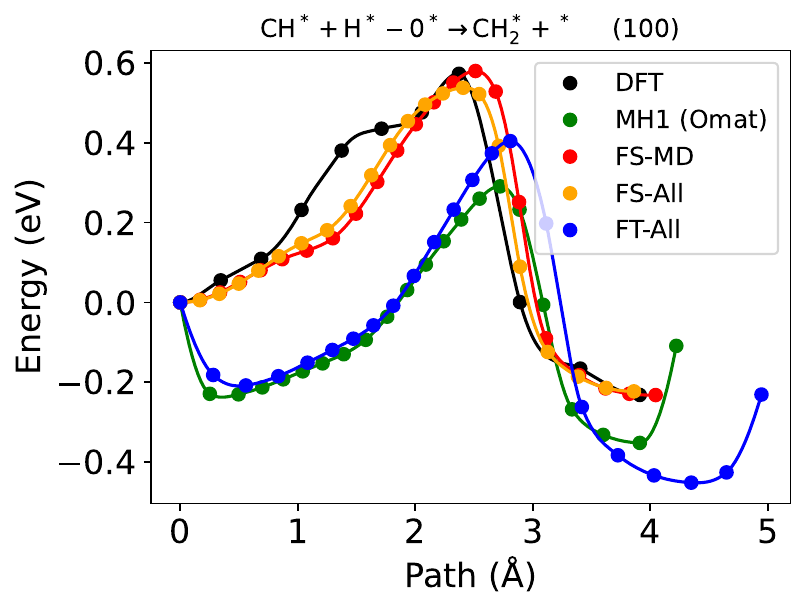}
\end{minipage}

\vspace{4pt}

\begin{minipage}[b]{0.48\textwidth}
    \centering
    \includegraphics[width=\linewidth,height=0.19\textheight,keepaspectratio]{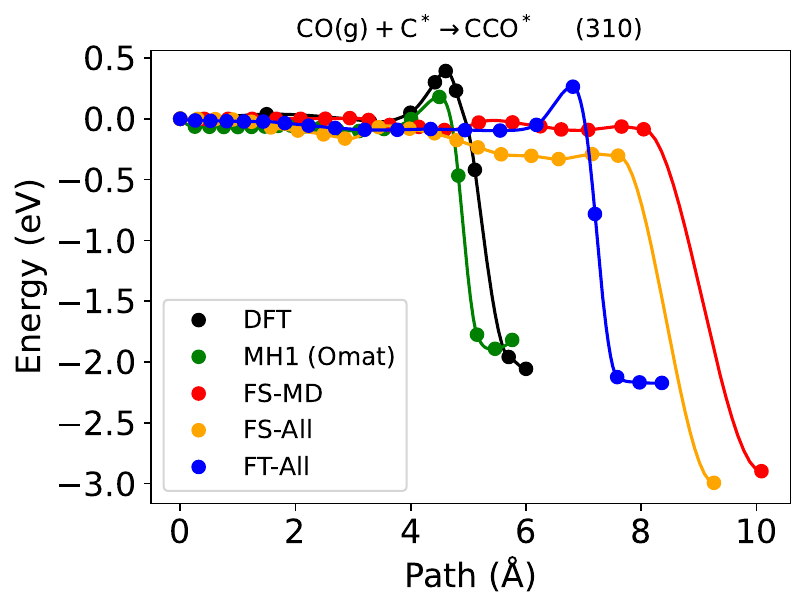}
\end{minipage}\hfill
\begin{minipage}[b]{0.48\textwidth}
    \centering
    \includegraphics[width=\linewidth,height=0.19\textheight,keepaspectratio]{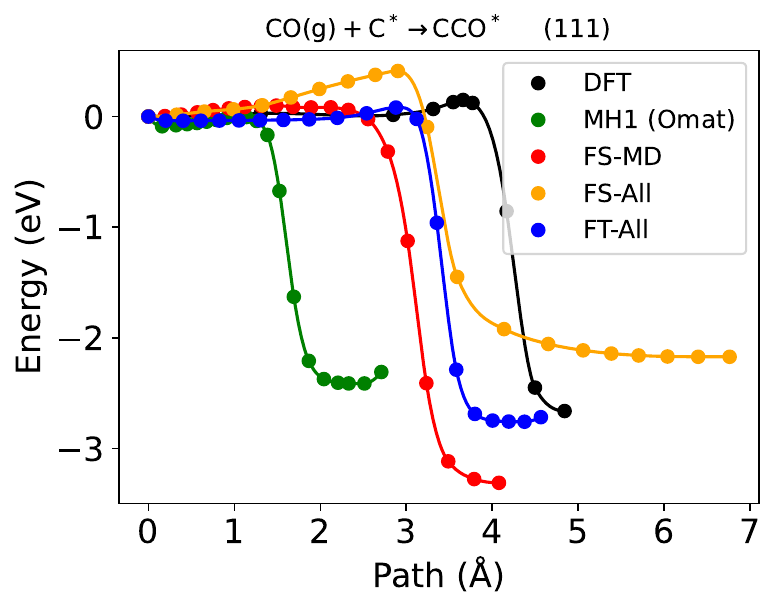}
\end{minipage}

\vspace{4pt}

\begin{minipage}[b]{0.48\textwidth}
    \centering
    \includegraphics[width=\linewidth,height=0.19\textheight,keepaspectratio]{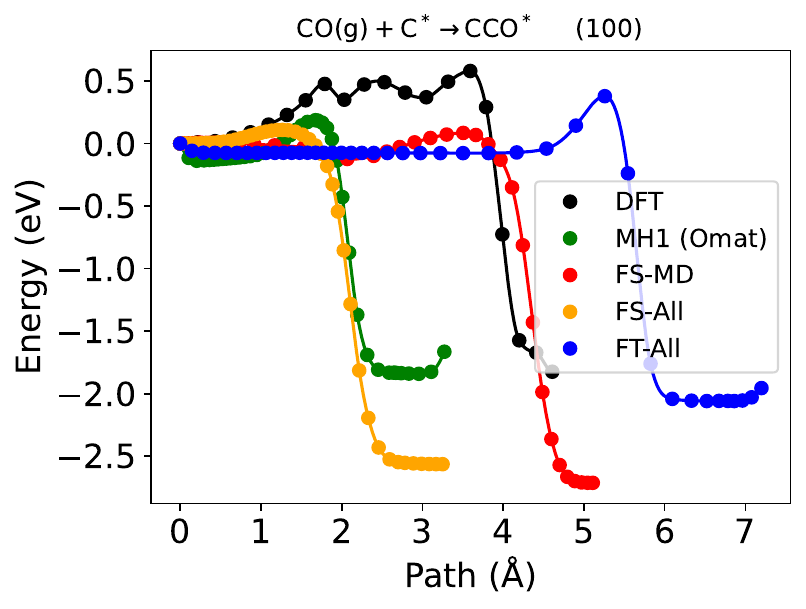}
\end{minipage}\hfill
\begin{minipage}[b]{0.48\textwidth}
    \centering
    \includegraphics[width=\linewidth,height=0.19\textheight,keepaspectratio]{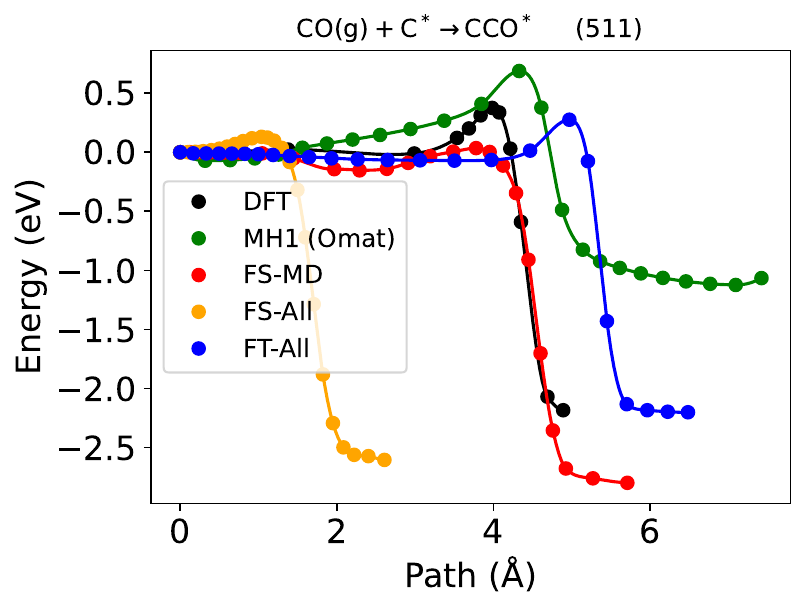}
\end{minipage}

\captionof{figure}{}
\label{fig:pengrole_c2_7fig}

\end{minipage}
\end{center}

    


    
    

\begin{minipage}[b]{0.48\textwidth}
    \centering
    \includegraphics[width=\linewidth,height=0.21\textheight,keepaspectratio]{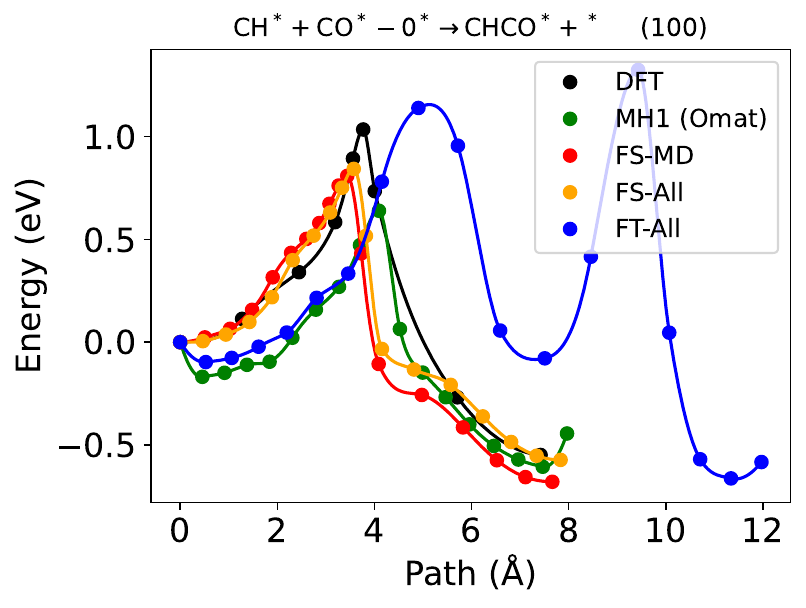}
\end{minipage}\hfill
\begin{minipage}[b]{0.48\textwidth}
    \centering
    \includegraphics[width=\linewidth,height=0.21\textheight,keepaspectratio]{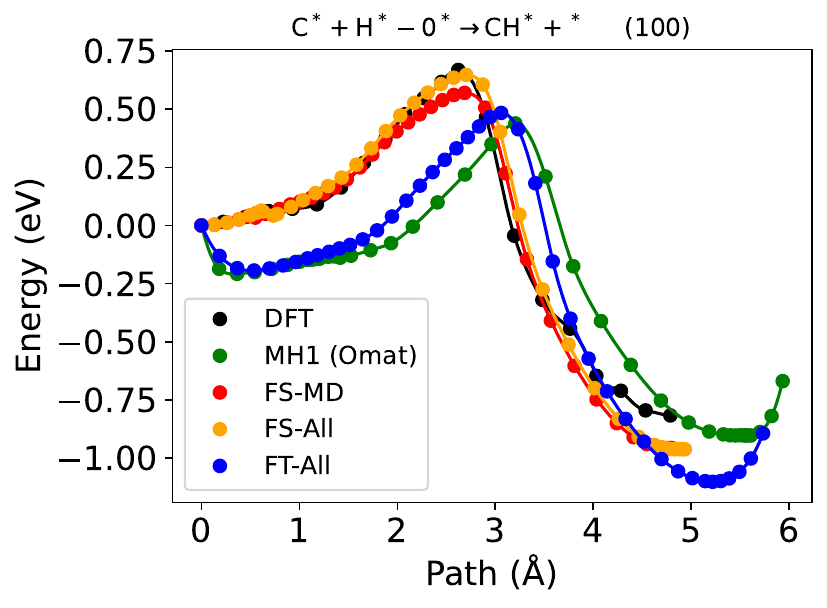}
\end{minipage}
\captionof{figure}{}
\label{fig:pengrole_c2_7fig}

\clearpage
\onecolumn

\subsection{\ce{CO2} reduction to \ce{C3} products \cite{tang_exp}}

\begin{center}
\begin{minipage}{0.95\textwidth}
\centering

\begin{minipage}[b]{0.44\textwidth}
    \centering
    \includegraphics[width=\linewidth,height=0.21\textheight,keepaspectratio]{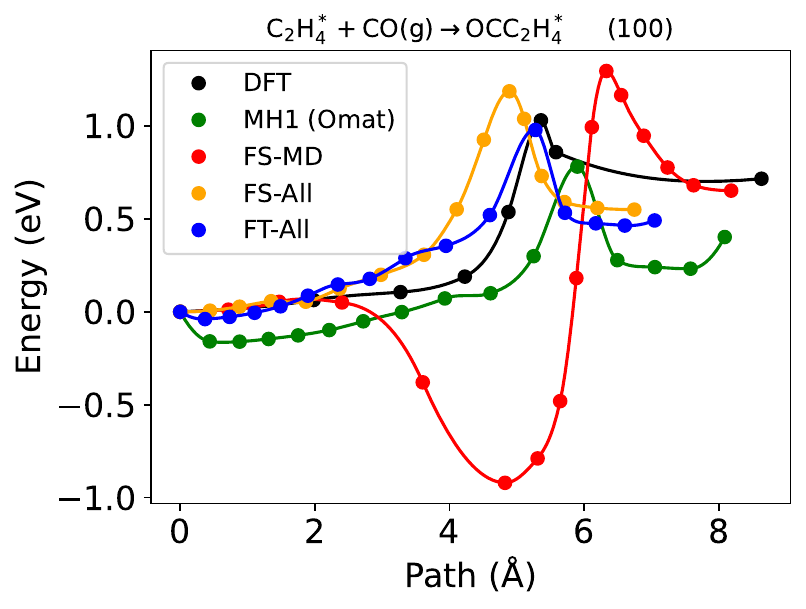}
\end{minipage}\hfill
\begin{minipage}[b]{0.44\textwidth}
    \centering
    \includegraphics[width=\linewidth,height=0.21\textheight,keepaspectratio]{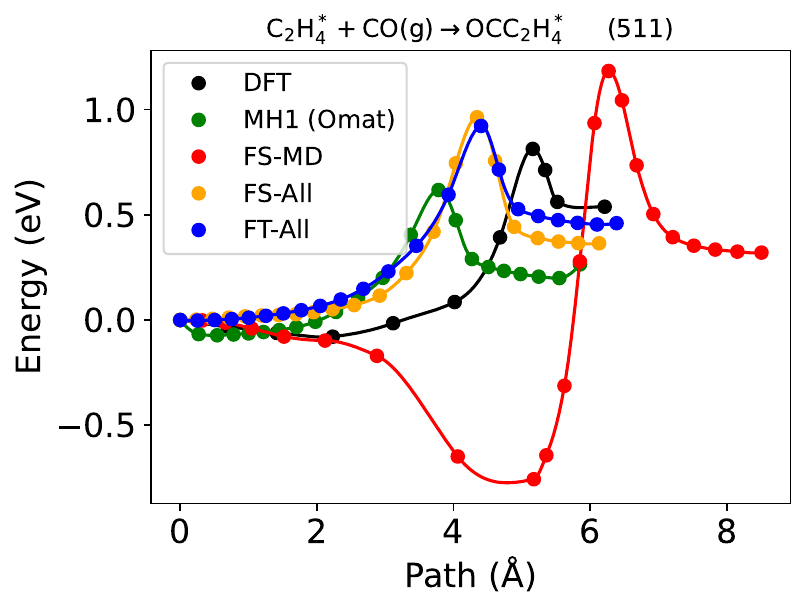}
\end{minipage}

\vspace{4pt}

\begin{minipage}[b]{0.44\textwidth}
    \centering
    \includegraphics[width=\linewidth,height=0.21\textheight,keepaspectratio]{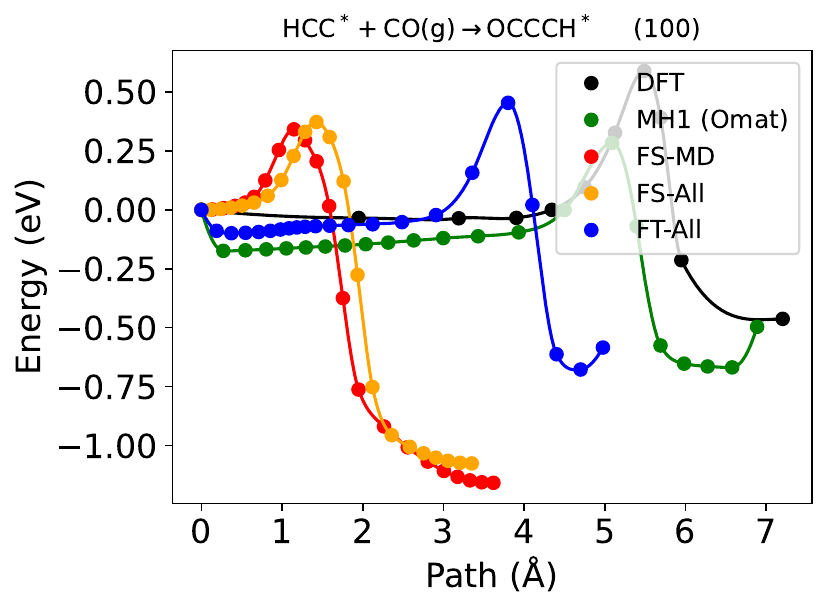}
\end{minipage}\hfill
\begin{minipage}[b]{0.44\textwidth}
    \centering
    \includegraphics[width=\linewidth,height=0.21\textheight,keepaspectratio]{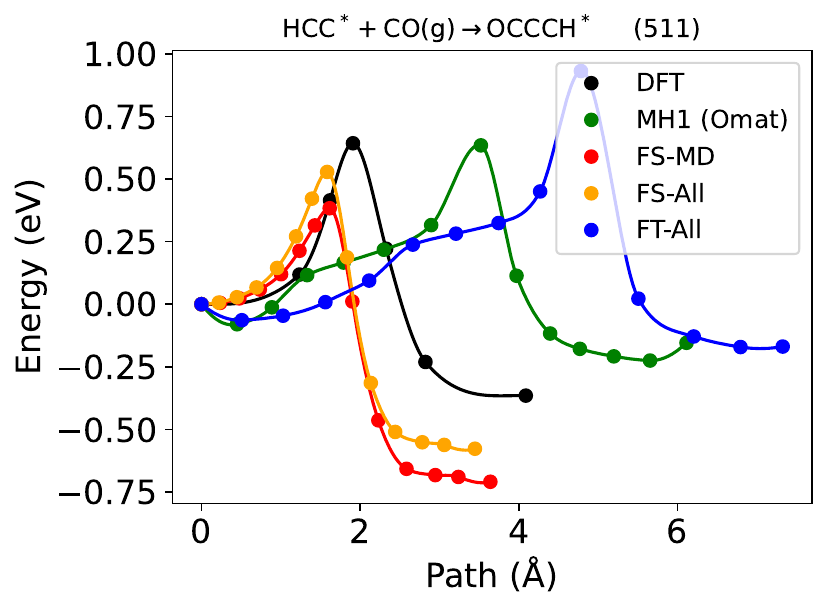}
\end{minipage}

\vspace{4pt}

\begin{minipage}[b]{0.44\textwidth}
    \centering
    \includegraphics[width=\linewidth,height=0.21\textheight,keepaspectratio]{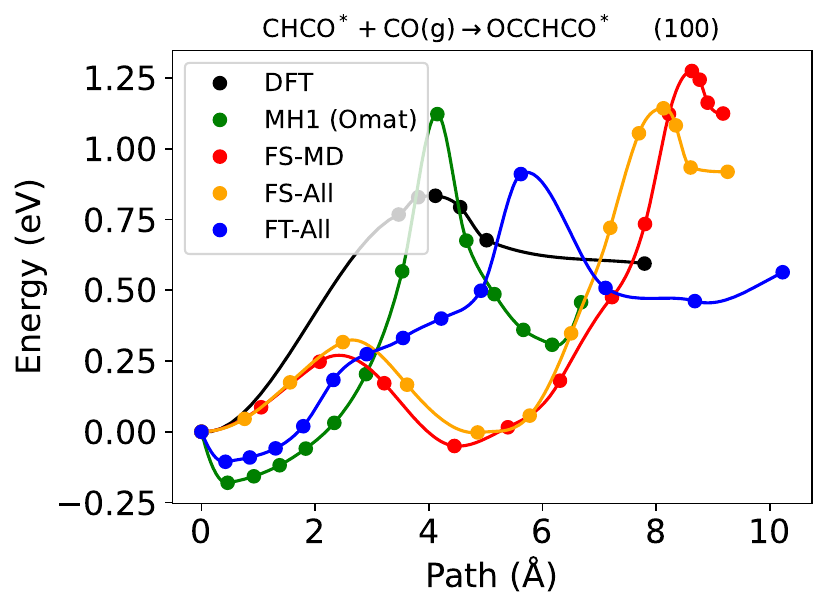}
\end{minipage}\hfill
\begin{minipage}[b]{0.44\textwidth}
    \centering
    \includegraphics[width=\linewidth,height=0.21\textheight,keepaspectratio]{si_figures/NEBs/TangExploring2021_4.pdf}
\end{minipage}

\vspace{4pt}

\begin{minipage}[b]{0.44\textwidth}
    \centering
    \includegraphics[width=\linewidth,height=0.21\textheight,keepaspectratio]{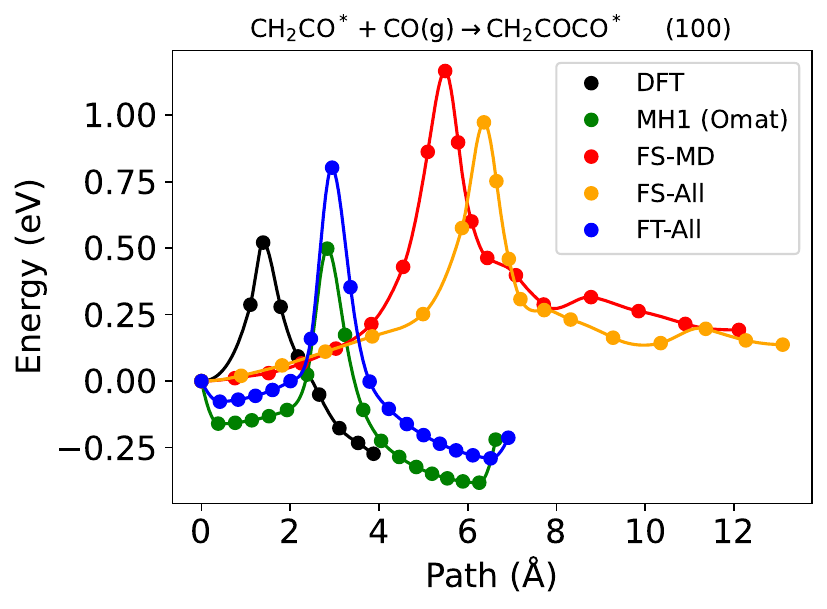}
\end{minipage}\hfill
\begin{minipage}[b]{0.44\textwidth}
    \centering
    \includegraphics[width=\linewidth,height=0.21\textheight,keepaspectratio]{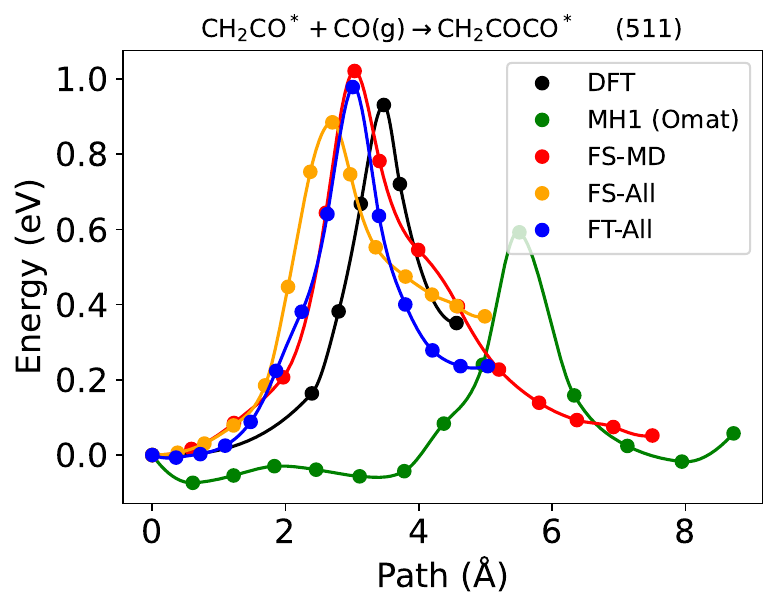}
\end{minipage}

\captionof{figure}{}
\label{fig:tang_c3_products}

\end{minipage}
\end{center}

\clearpage
\onecolumn

\begin{center}
\begin{minipage}{0.95\textwidth}
\centering

\begin{minipage}[b]{0.44\textwidth}
    \centering
    \includegraphics[width=\linewidth,height=0.21\textheight,keepaspectratio]{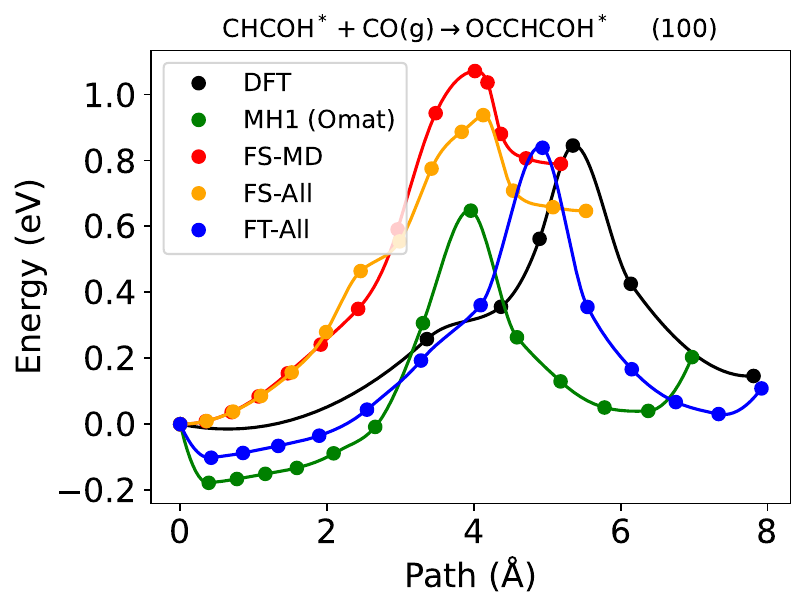}
\end{minipage}\hfill
\begin{minipage}[b]{0.44\textwidth}
    \centering
    \includegraphics[width=\linewidth,height=0.21\textheight,keepaspectratio]{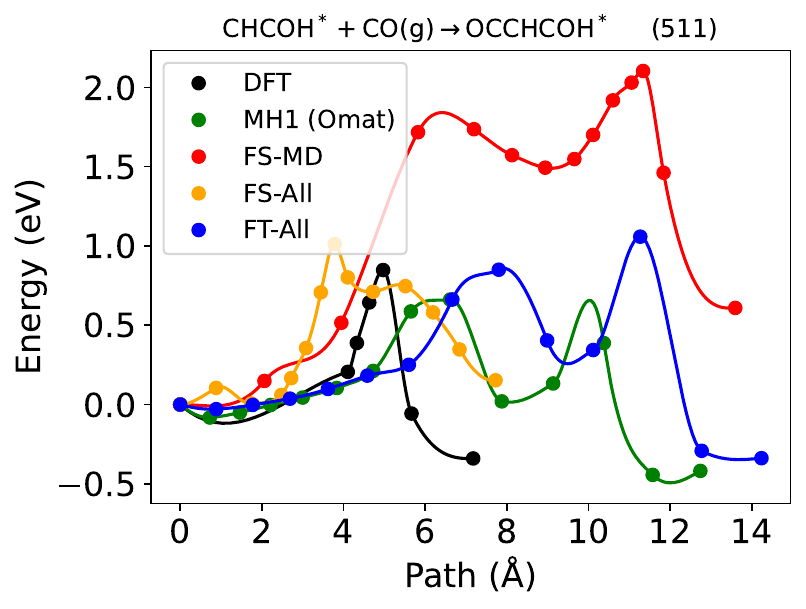}
\end{minipage}

\vspace{4pt}

\begin{minipage}[b]{0.44\textwidth}
    \centering
    \includegraphics[width=\linewidth,height=0.21\textheight,keepaspectratio]{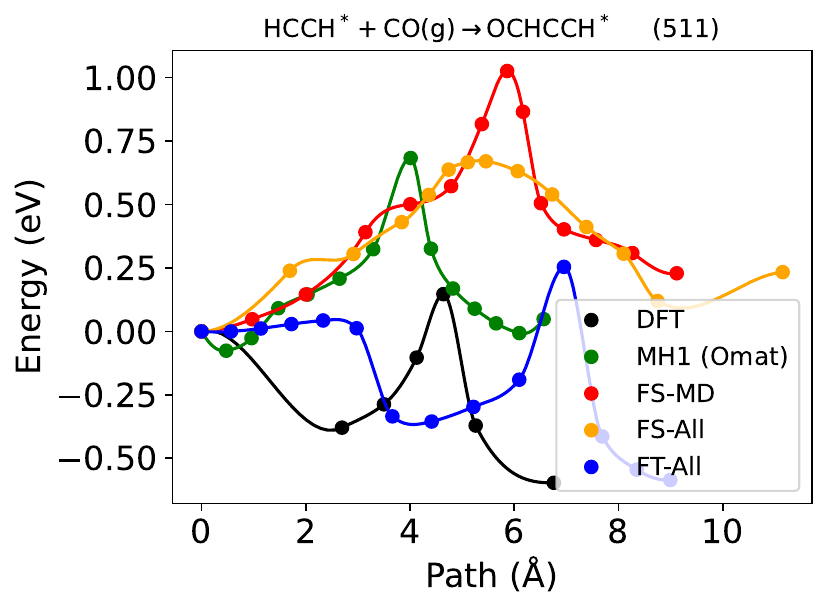}
\end{minipage}\hfill
\begin{minipage}[b]{0.44\textwidth}
    \centering
    \includegraphics[width=\linewidth,height=0.21\textheight,keepaspectratio]{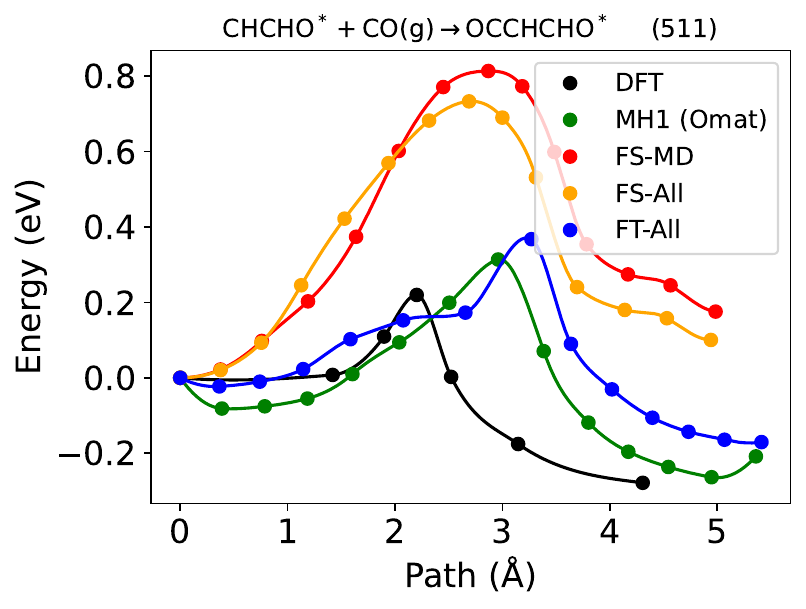}
\end{minipage}

\vspace{4pt}

\begin{minipage}[b]{0.44\textwidth}
    \centering
    \includegraphics[width=\linewidth,height=0.21\textheight,keepaspectratio]{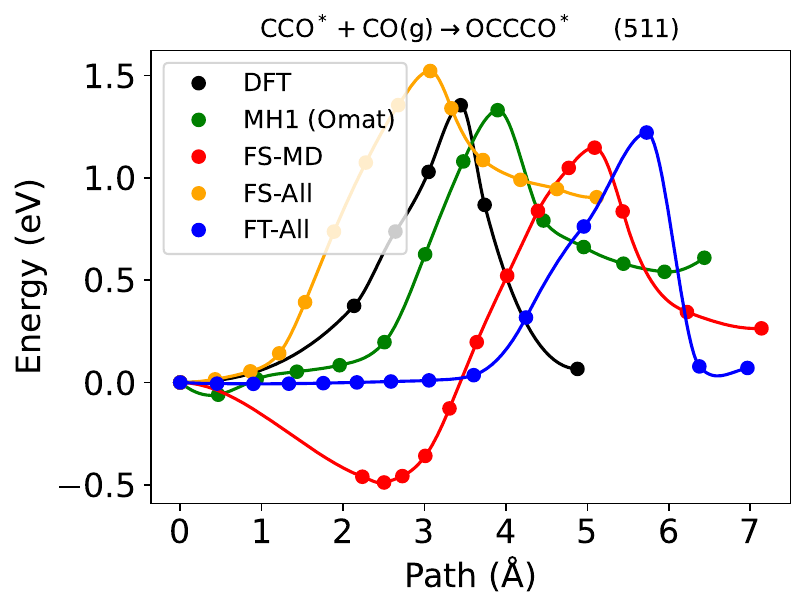}
\end{minipage}\hfill
\begin{minipage}[b]{0.44\textwidth}
    \centering
    \includegraphics[width=\linewidth,height=0.21\textheight,keepaspectratio]{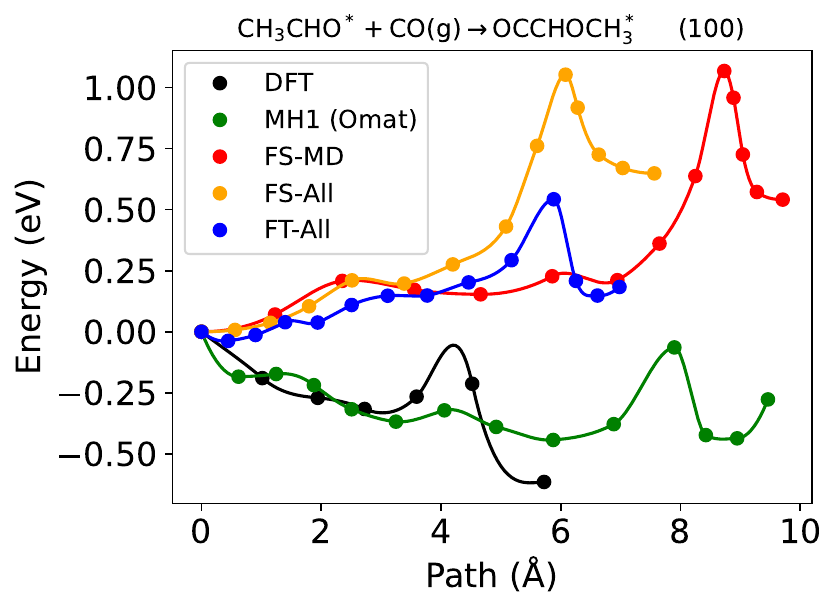}
\end{minipage}

\captionof{figure}{}
\label{fig:tang_6fig}

\end{minipage}
\end{center}

\twocolumn

\clearpage
\onecolumn

\begin{center}
\begin{minipage}{0.95\textwidth}
\centering

\begin{minipage}[b]{0.44\textwidth}
    \centering
    \includegraphics[width=\linewidth,height=0.21\textheight,keepaspectratio]{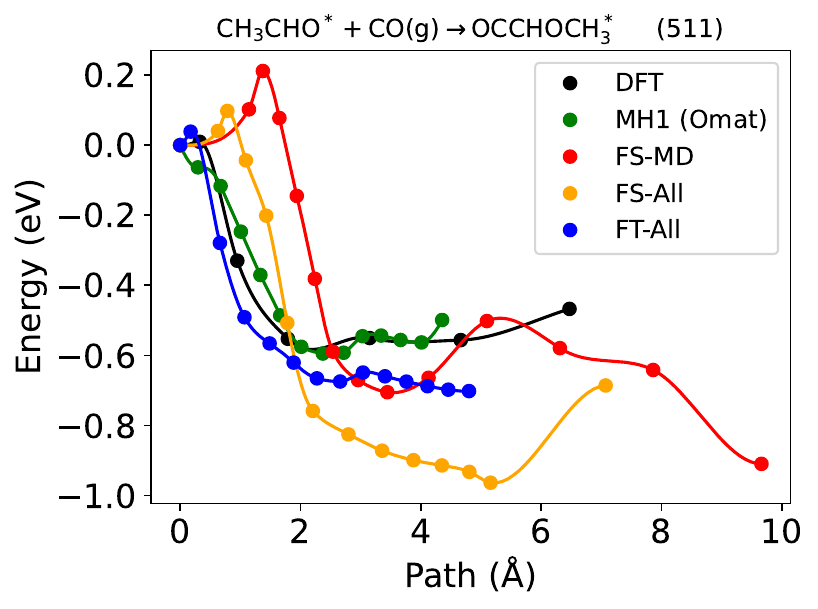}
\end{minipage}\hfill
\begin{minipage}[b]{0.44\textwidth}
    \centering
    \includegraphics[width=\linewidth,height=0.21\textheight,keepaspectratio]{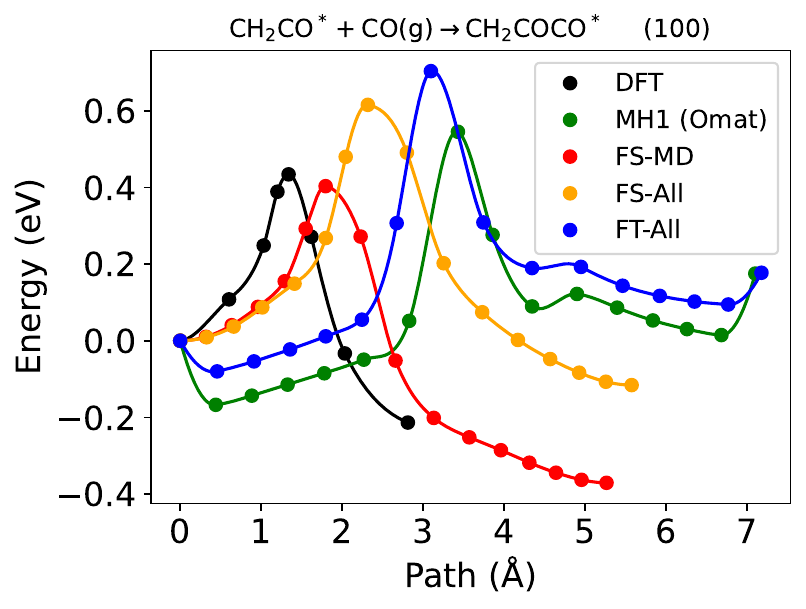}
\end{minipage}

\vspace{4pt}

\begin{minipage}[b]{0.44\textwidth}
    \centering
    \includegraphics[width=\linewidth,height=0.21\textheight,keepaspectratio]{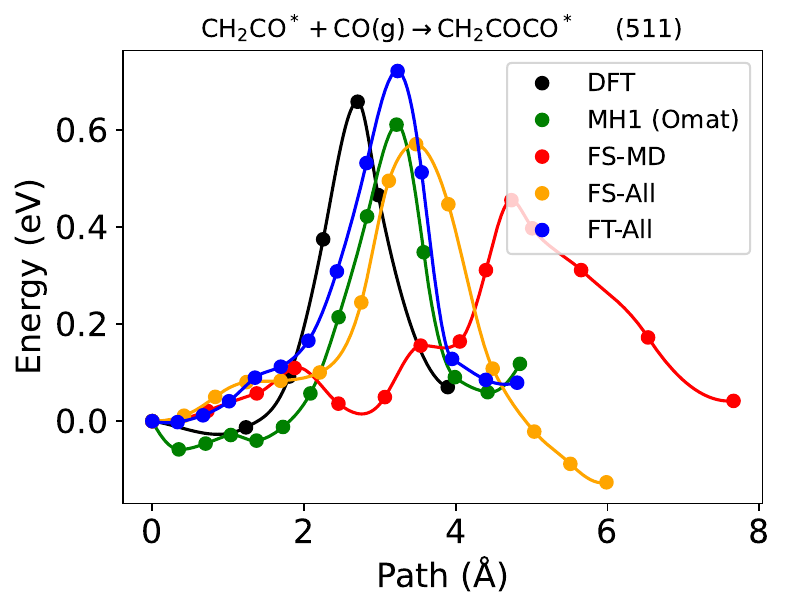}
\end{minipage}\hfill
\begin{minipage}[b]{0.44\textwidth}
    \centering
    \includegraphics[width=\linewidth,height=0.21\textheight,keepaspectratio]{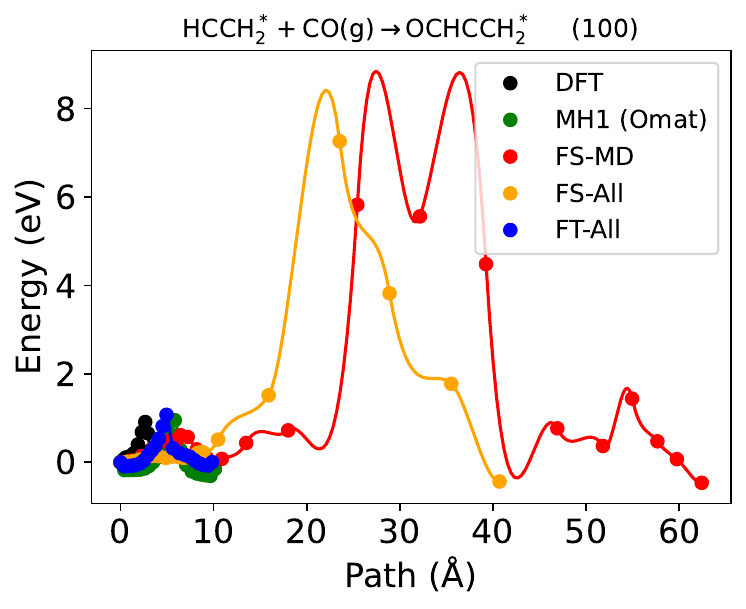}
\end{minipage}

\vspace{4pt}

\begin{minipage}[b]{0.44\textwidth}
    \centering
    \includegraphics[width=\linewidth,height=0.21\textheight,keepaspectratio]{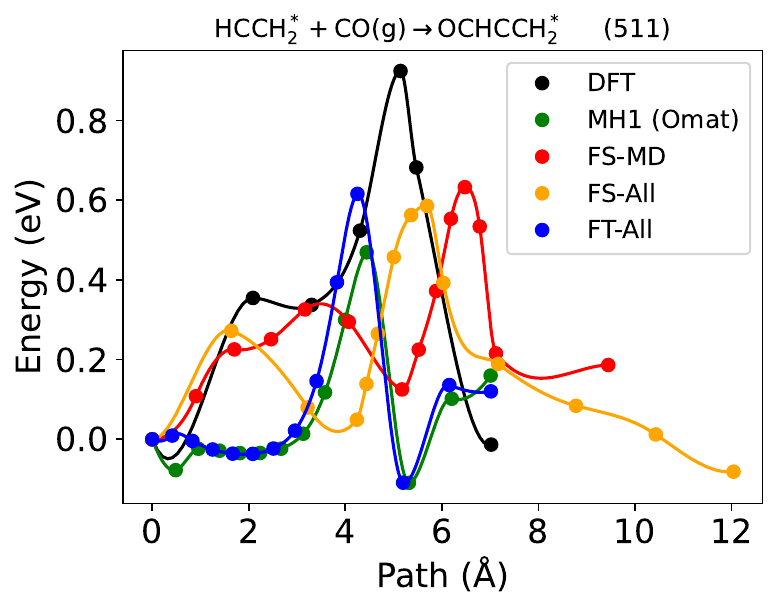}
\end{minipage}\hfill
\begin{minipage}[b]{0.44\textwidth}
    \centering
    \includegraphics[width=\linewidth,height=0.21\textheight,keepaspectratio]{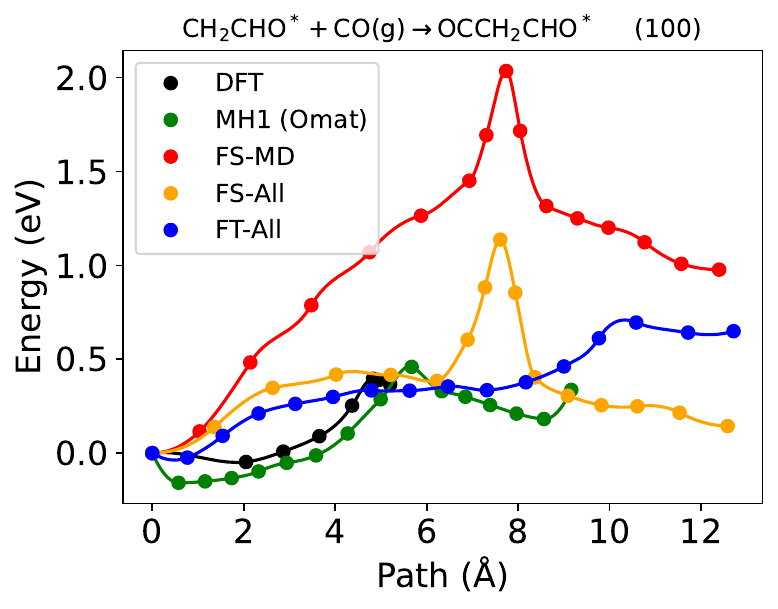}
\end{minipage}

\vspace{4pt}

\begin{minipage}[b]{0.44\textwidth}
    \centering
    \includegraphics[width=\linewidth,height=0.21\textheight,keepaspectratio]{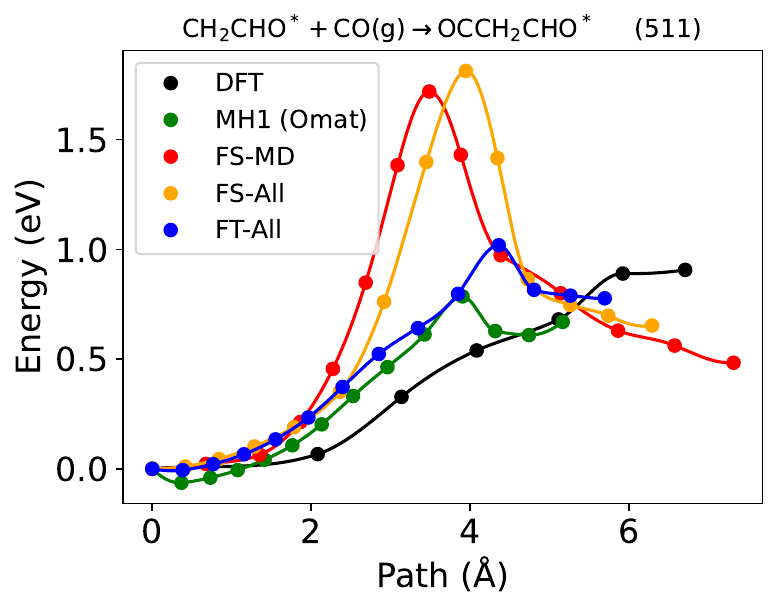}
\end{minipage}\hfill
\begin{minipage}[b]{0.44\textwidth}
    \centering
    \includegraphics[width=\linewidth,height=0.21\textheight,keepaspectratio]{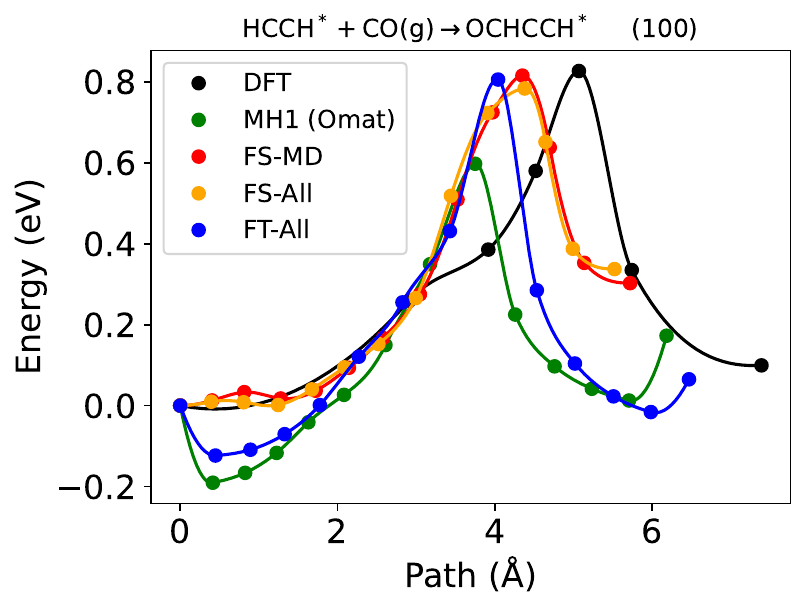}
\end{minipage}

\captionof{figure}{}
\label{fig:tang_8grid_b}

\end{minipage}
\end{center}

\clearpage
\clearpage
\onecolumn

\subsection{Tafel step in HER \cite{test_set_tang_modeling}}

\begin{center}
\begin{minipage}{0.98\textwidth}
\centering

\begin{minipage}[b]{0.48\textwidth}
    \centering
    \includegraphics[width=\linewidth,height=0.21\textheight,keepaspectratio]{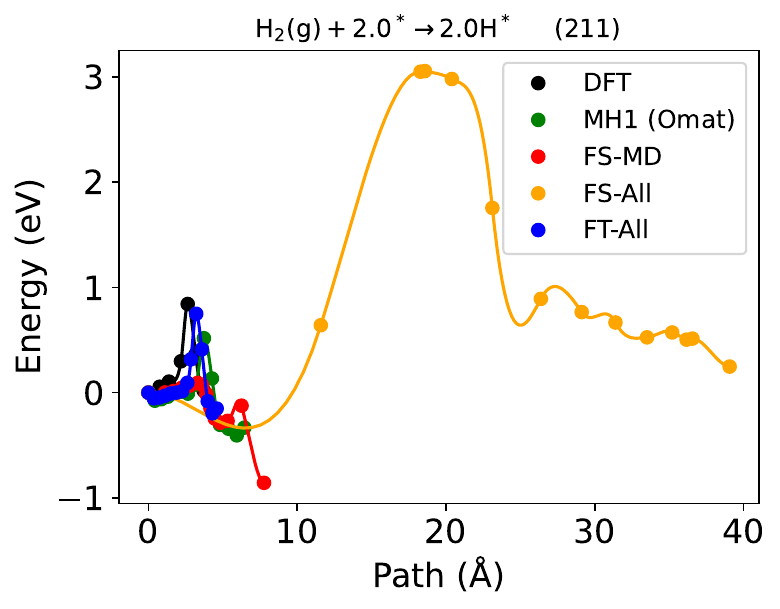}
\end{minipage}\hfill
\begin{minipage}[b]{0.48\textwidth}
    \centering
    \includegraphics[width=\linewidth,height=0.21\textheight,keepaspectratio]{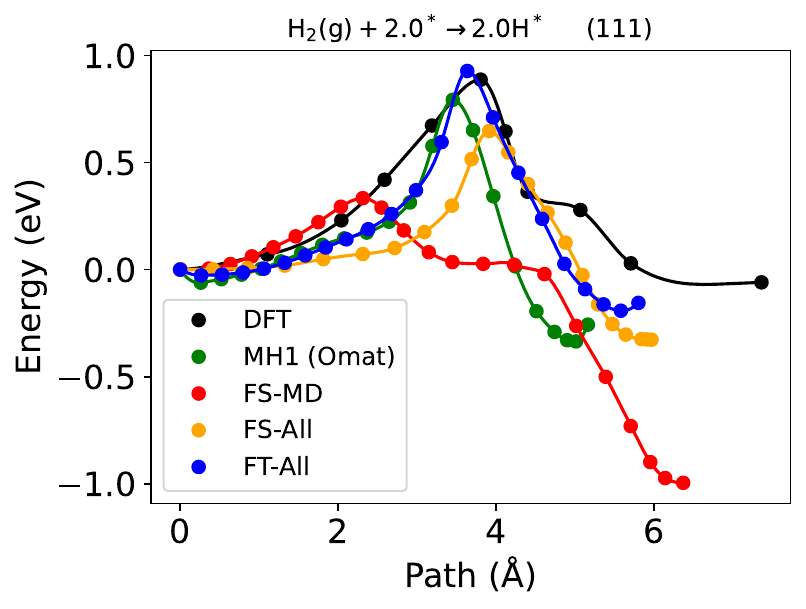}
\end{minipage}

\vspace{4pt}

\begin{minipage}[b]{0.48\textwidth}
    \centering
    \includegraphics[width=\linewidth,height=0.21\textheight,keepaspectratio]{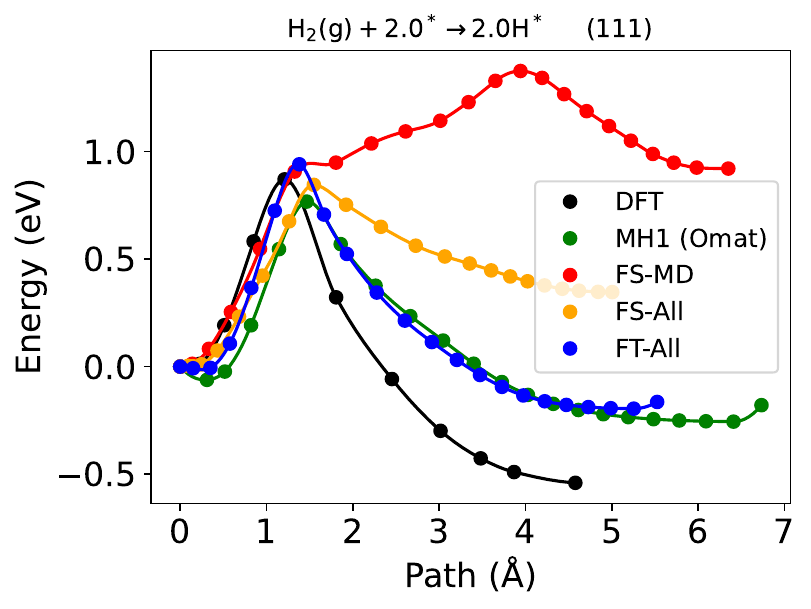}
\end{minipage}\hfill
\begin{minipage}[b]{0.48\textwidth}
    \centering
    \includegraphics[width=\linewidth,height=0.21\textheight,keepaspectratio]{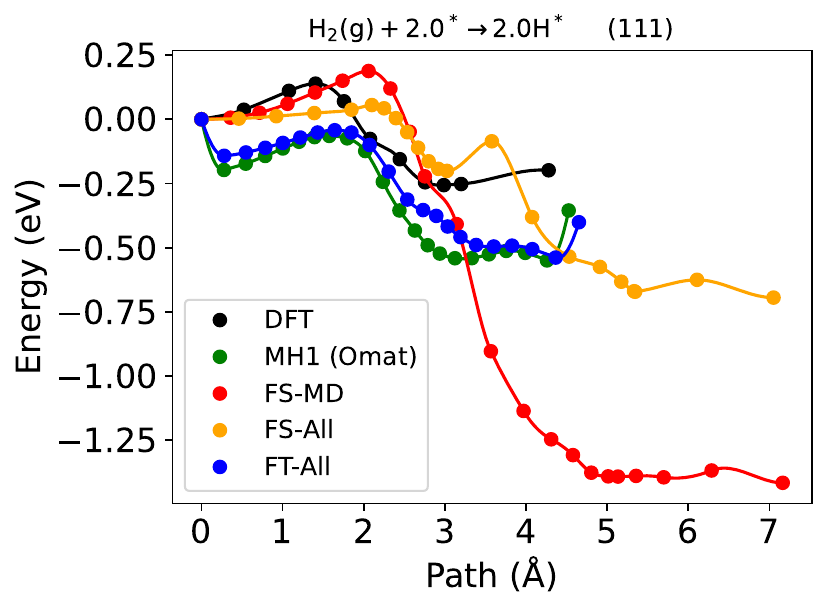}
\end{minipage}

\vspace{4pt}

\begin{minipage}[b]{0.48\textwidth}
    \centering
    \includegraphics[width=\linewidth,height=0.21\textheight,keepaspectratio]{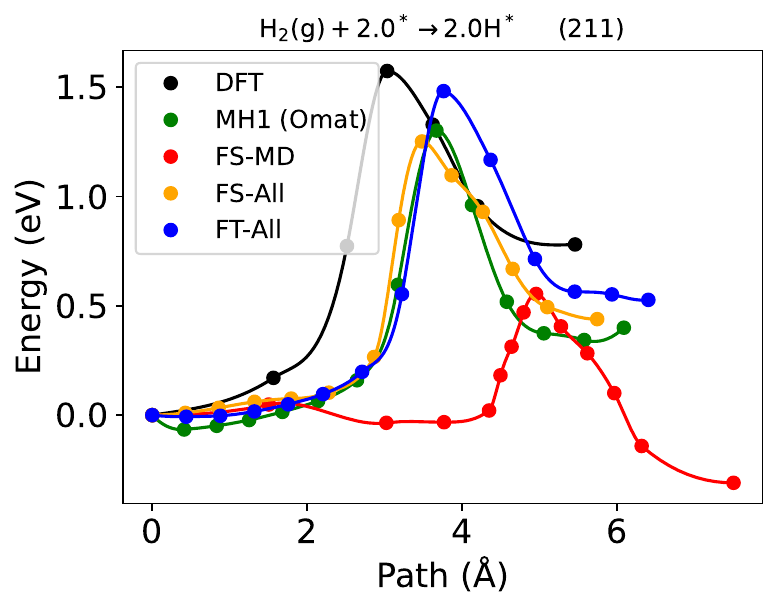}
\end{minipage}\hfill
\begin{minipage}[b]{0.48\textwidth}
    \centering
    \includegraphics[width=\linewidth,height=0.21\textheight,keepaspectratio]{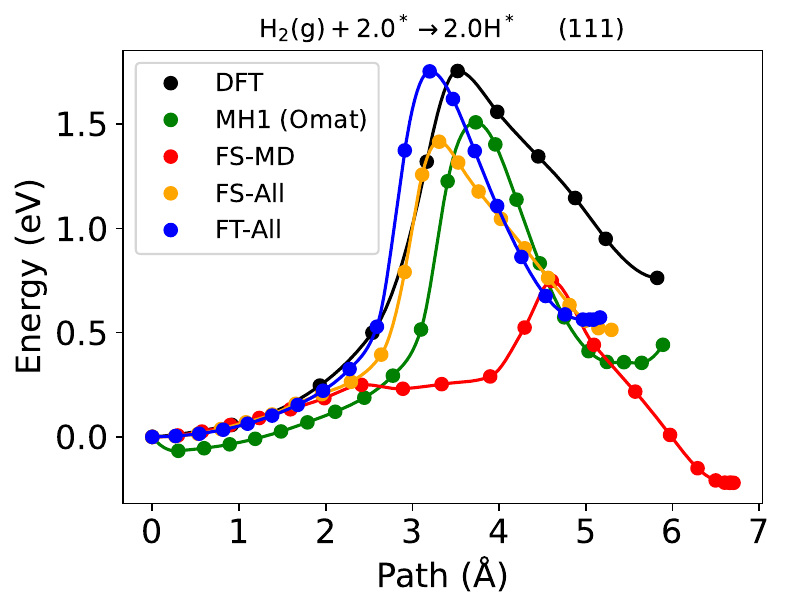}
\end{minipage}

\vspace{4pt}

\begin{minipage}[b]{0.48\textwidth}
    \centering
    \includegraphics[width=\linewidth,height=0.21\textheight,keepaspectratio]{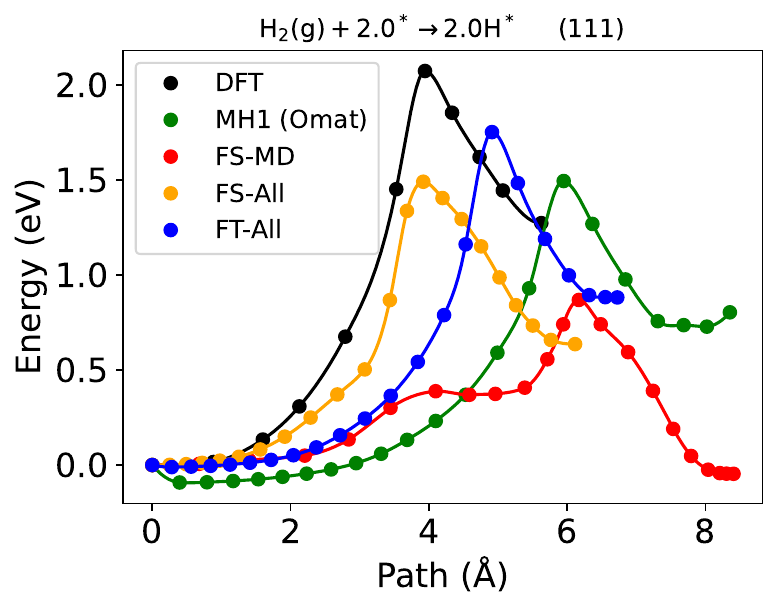}
\end{minipage}\hfill
\begin{minipage}[b]{0.48\textwidth}
    \centering
    \includegraphics[width=\linewidth,height=0.21\textheight,keepaspectratio]{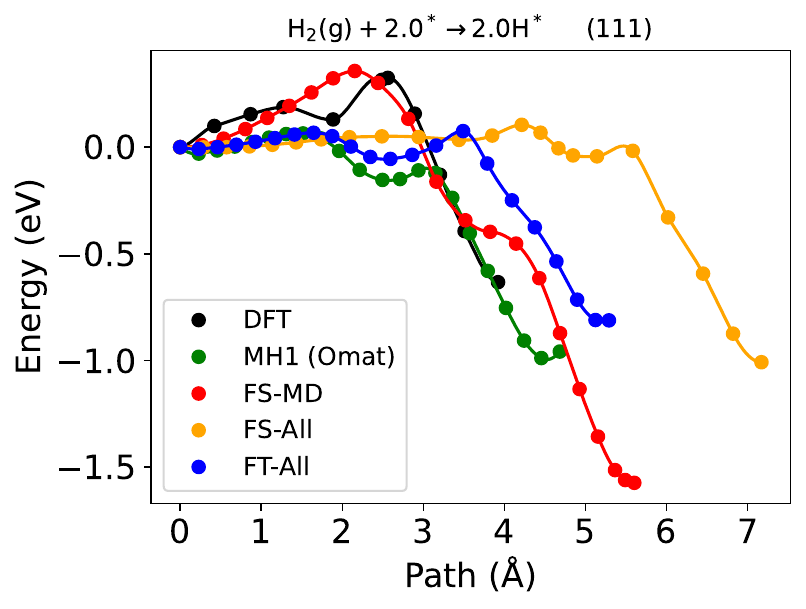}
\end{minipage}

\captionof{figure}{}
\label{fig:tafel_her_8fig}

\end{minipage}
\end{center}

\clearpage
\onecolumn

\begin{center}
\begin{minipage}{0.98\textwidth}
\centering

\begin{minipage}[b]{0.48\textwidth}
    \centering
    \includegraphics[width=\linewidth,height=0.21\textheight,keepaspectratio]{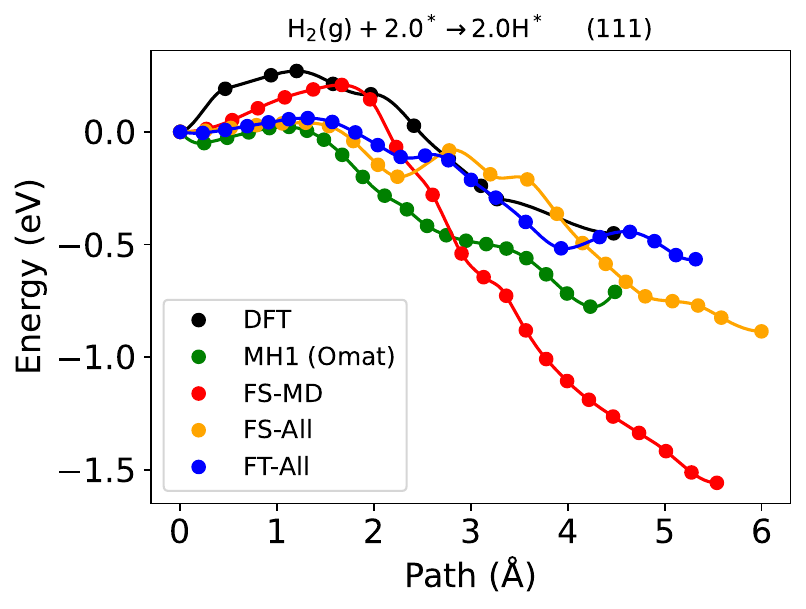}
\end{minipage}\hfill
\begin{minipage}[b]{0.48\textwidth}
    \centering
    \includegraphics[width=\linewidth,height=0.21\textheight,keepaspectratio]{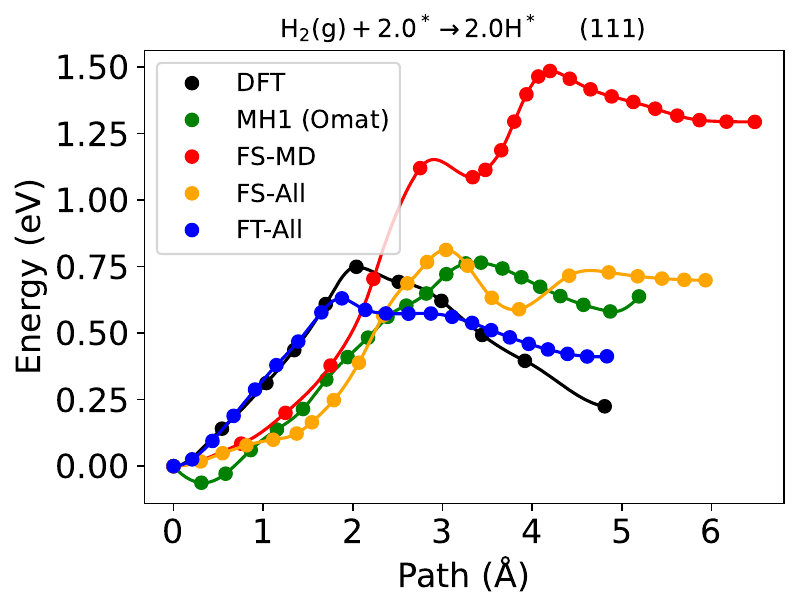}
\end{minipage}

\vspace{4pt}

\begin{minipage}[b]{0.48\textwidth}
    \centering
    \includegraphics[width=\linewidth,height=0.21\textheight,keepaspectratio]{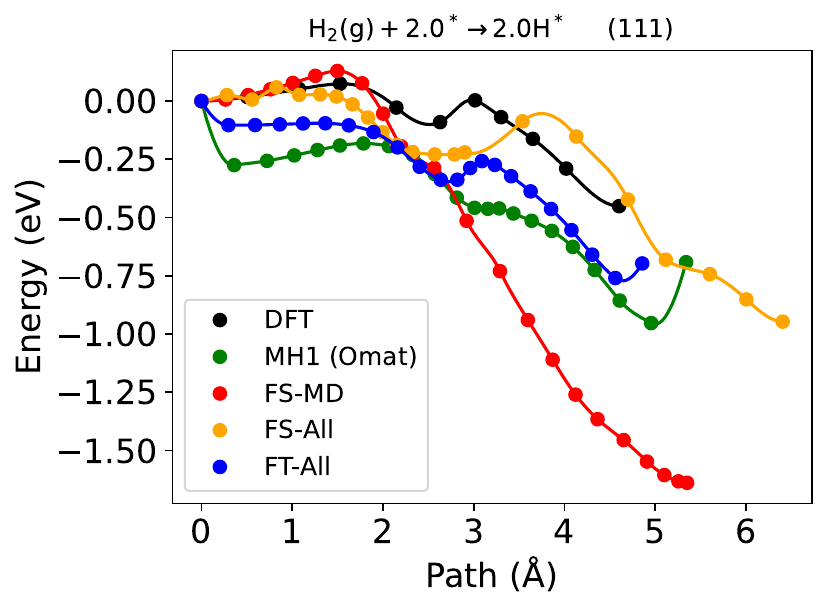}
\end{minipage}\hfill
\begin{minipage}[b]{0.48\textwidth}
    \centering
    \includegraphics[width=\linewidth,height=0.21\textheight,keepaspectratio]{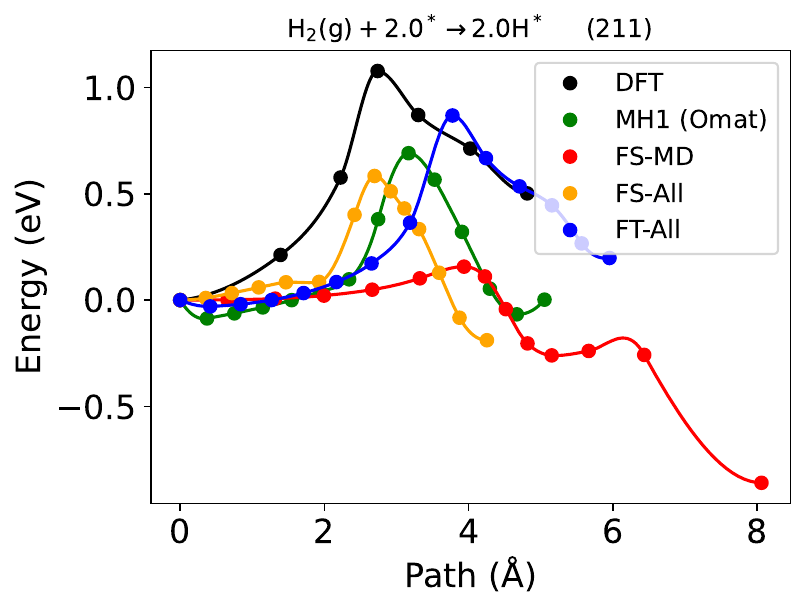}
\end{minipage}

\vspace{4pt}

\begin{minipage}[b]{0.48\textwidth}
    \centering
    \includegraphics[width=\linewidth,height=0.21\textheight,keepaspectratio]{si_figures/NEBs/TangModeling2020_105.pdf}
\end{minipage}

\captionof{figure}{}
\label{fig:tangmodeling_5fig}

\end{minipage}
\end{center}

\clearpage
\section{MLIP applied to different transition states}

We have applied the FT-All model to the reaction of \ce{CHCOH}* \ce{+ CO(g) -> OCCHCOH}* on Cu(001)
 that were previously done with different MLIPs or DFT calcualtors. For each case, the FT-All is applied as a single-point calculation at the converged NEB pathways. These results shows that with model, one can achieve a different transition states, which with the same calculator can have up to 0.3 eV difference in energy.
\begin{figure}[H]
    \centering  
\includegraphics[width=0.5\textwidth]{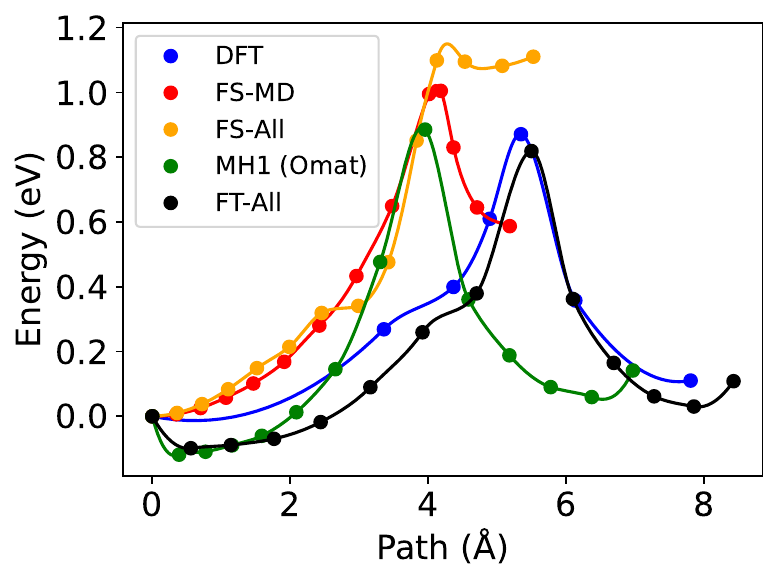}
    \caption{}
\end{figure}
\clearpage
\section{NEB convergence for each FT MLIP}
\begin{table*}[h!]
\centering

\begin{tabular}{ccc}
\toprule
 \textbf{Model} & \textbf{\reacbarrier{}}  & \textbf{Convergence}   \\  

\hline
 FT-BMA & 0.141  & 100 \%    \\  
 FT-MD(M)  & 0.159  & 91 \%   \\   
 FT-MD(MO)  & 0.149  & 91 \%   \\   
 FT-CE  & 0.143  & 100 \%   \\   
 FT-BMA+MD(M)  & 0.153  & 100 \%   \\   
 FT-All  & 0.142  & 100 \%   \\   \bottomrule

 \end{tabular}

\caption{\reacbarrier{} MAEs compared against DFT energies for our different models predicting the 23 reactions in Ref. \cite{tang_exp}. The convergence column shows the percentage of NEBs that converged from the 23 reactions.}
\label{table:neb_co2rr}
\end{table*}

\clearpage
\section{Screening adsorption energies on binary alloys}

The list of 33 adsorbates that were targeted in the screening task.
\begin{table}[h!]
\centering
\begin{tabular}{cccc}
\hline
\multicolumn{4}{c}{\textbf{Adsorbates}} \\
\hline
$\mathrm{^*C^*C}$ & $\mathrm{^*CCH}$ & $\mathrm{^*CCH_2}$ & $\mathrm{^*CCH_3}$ \\
$\mathrm{^*CCHO}$ & $\mathrm{^*CCHOH}$ & $\mathrm{^*CCO}$ & $\mathrm{^*CH^*CH}$ \\
$\mathrm{^*CH_2}$ & $\mathrm{^*CH_2^*O}$ & $\mathrm{^*CH_3}$ & $\mathrm{^*CH_4}$ \\
$\mathrm{^*CHCH_2}$ & $\mathrm{^*CHCHO}$ & $\mathrm{^*CHCO}$ & $\mathrm{^*CHO}$ \\
$\mathrm{^*CHO^*CHO}$ & $\mathrm{^*CHOH}$ & $\mathrm{^*COCH_2O}$ & $\mathrm{^*COCH_3}$ \\
$\mathrm{^*COCHO}$ & $\mathrm{^*COH}$ & $\mathrm{^*COHCHO}$ & $\mathrm{^*COHCOH}$ \\
$\mathrm{^*NH_2}$ & $\mathrm{^*NH_3}$ & $\mathrm{^*NONH}$ & $\mathrm{^*OCHCH_2}$ \\
$\mathrm{^*OHCH_3}$ & $\mathrm{^*OOH}$ & $\mathrm{CH^*COH}$ & $\mathrm{CH_2^*CO}$ \\
$\mathrm{CO^*COH}$ &  &  &  \\
\hline
\end{tabular}
\caption{Adsorbed species with surface notation ($^*$) for the binding atom}
\end{table}

\clearpage

\bibliography{sn-bibliography}

\end{document}